\documentclass[twocolumn,showpacs,showkeys,preprintnumbers,amsmath,amssymb]{revtex4}

\usepackage{dcolumn}
\usepackage{bm}
\usepackage{physics}
\usepackage{graphicx,color}
\usepackage{bbm}
\usepackage{amsmath,amssymb,mathrsfs}
\usepackage{url}
\usepackage{hyperref}

\def\>{\rangle}
\def\<{\langle}

\newcommand{\map}[1]{\mathcal{#1}}

\newcommand{\hS}{{\widehat {\map S} }}

\usepackage{qcircuit}

\begin{document}



\title{The quantum network as an environment}

\author{Erik Aurell}
\email{eaurell@kth.se}
\affiliation{KTH -- Royal Institute of Technology, AlbaNova University Center, SE-106 91 Stockholm, Sweden}%

\author{Roberto Mulet}
\email{roberto.mulet@gmail.com}
\affiliation{
Group of Complex Systems and Statistical Physics. Department of Theoretical Physics,
Physics Faculty, University of Havana, Cuba}

\author{Jan Tuziemski}%
\email{jan.tuziemski@fysik.su.se}
\affiliation{Department of Physics, Stockholm University, AlbaNova University Center, Stockholm SE-106 91 Sweden}
\affiliation{Nordita, Royal Institute of Technology and Stockholm University,Roslagstullsbacken 23, SE-106 91 Stockholm, Sweden}
\affiliation{ Department of Applied Physics and Mathematics, Gdańsk University of Technology,
80-233 Gdańsk, Poland}
\altaffiliation[On leave from ]{ Gdańsk University of Technology}

\date{\today}

\begin{abstract}
A quantum system interacting with other quantum systems in a network experiences these
other systems as an effective environment. This environment is the result of integrating out
all the other degrees of freedom in the network, and can be represented by a Feynman-Vernon
influence functional (IF) acting on system of interest. 
A network is characterized by the constitutive systems, how they interact,
and the topology of those interactions.
Here we show that
for networks having the topology of locally tree-like graphs,
the Feynman-Vernon
influence functional can be determined in a new version of the
cavity or Belief Propagation (BP) method.
In the BP update stage, cavity IFs are
mapped to cavity IFs, while in the BP output stage
cavity IFs are combined to output IFs.
We compute the fixed point of of this version of BP for 
harmonic oscillator systems interacting uniformly.
We discuss Replica Symmetry and the effects of disorder in this context.
\end{abstract}

\maketitle

The cavity method is a way to simultaneously compute all marginals of a Gibbs-Boltzmann distribution when the interaction graph has no short loops. It is computationally efficient, taking a time polynomial in system size. It is exact when interactions form tree graphs, and in many cases asymptotically exact when the graph size tends to infinity. Comprehensive modern references are  \cite{MezardMontanari,richardson2008modern,WainwrightJordan2008}.
As introduced by Bethe~\cite{Bethe1935} the cavity method in statistical physics is a mean-field approximation; physical lattices in more than one dimension have many short loops. In other sciences the assumption of no short loops can be much more accurate and/or much more meaningful, and in engineered systems it can hold by design. This is the basis for the success of the cavity method in applications to biology, sociology and ICT, then more often called Belief Propagation \cite{YedidiaFreemanWeiss2003}.

A cavity method for quantum systems was introduced already in 1973~\cite{Abou_Chacra1973}. In that pioneering version, the object is the time-independent wave function on a Bethe lattice or a Cayley tree; cavity equations connect different sites where energy enters as a parameter. 
Recent applications 
of this family of methods have been to bosons on the Cayley tree~\cite{Dupont2020}, and to the Anderson transition on random graphs~\cite{Parisi2019,GarciaMata2020}. 
A second type of applications of the cavity method to quantum problems 
is to thermal equilibrium states~\cite{Laumann2008,Bapst2013}. 
A selection of other papers addressing quantum cavity method through different approaches are
\cite{Hastings2007,Leifer2008,PoulinBilgin2008,IoffeMezard2010,DimitrovaMezard2011,LoeligerVontobel2017,Renes2017}.

Here we introduce a new form of quantum cavity to describe the evolution of 
a quantum system in real time. As the essence of the cavity method is to marginalize
to the variables of interest, the real-time quantum cavity is an open
system, and the basic object of study is the reduced density matrix of one system. In contrast to quantum cavity for thermal equilibrium states, each system is described by \textit{two} histories
and \textit{two} Feynman path integrals. The result of integrating out all the other histories in the network is an influence functional acting
on the system of interest only~\cite{FeynmanVernon1963}.
The goal of this paper is to discuss the properties and meaning of these influence
functionals, and show that they can be computed in closed form in some simple 
but still interesting cases.
The overall message is that a quantum network behaves as quantum environment with computable
properties. 


Many current and projected quantum computing architectures are based on units with limited connectivity,
and properties of networks of such systems have been investigated widely \cite{Beals2013,Boixo2018,Cross2019,NamMaslov2019}.
The quantum annealing method has been applied to solve (classically) hard combinatorial optimization problems of such topologies~\cite{DasChakrabarti2008,Bapst2013}. The real-time quantum cavity method introduced here gives
a new approach to describe dissipation and decoherence of
the states involved in such algorithms.
We will outline the relevance of Replica Symmetry and quenched disorder
to these problems.

\section{The real-time quantum cavity and Replica Symmetry}
\label{sec:theproblem}
Consider a network of quantum systems denoted by $X_1 \ldots X_N$, the interaction pattern of which has a locally tree-like structure. 
The dynamics of the network $\rho(t) =  U(t) \rho_0 U^{\dagger}(t)$ can be expressed in the path-integral form
\begin{widetext}

\begin{eqnarray}
U(t) \cdot  U^{\dagger}(t) &=& \int D X_i D Y_i \cdots \exp \frac{i}{\hbar} \left(S\left[X_i \right] +S\left[X_j \right] -S\left[Y_i \right] - S\left[Y_j \right] + S\left[X_i,X_j \right] -S\left[Y_i,Y_j \right]  + \cdots    \right) \nonumber \\
\label{eq:evolution}
&& \qquad\qquad \rho_0\left(x_i(t_i),\ldots,y_i(t_i),\ldots\right)
\end{eqnarray}
\end{widetext}
where $\rho_0\left(x_i(t_i),\ldots,y_i(t_i),\ldots\right)$
is the initial density operator of 
the whole network. 
For each system there is one "forward path" (denoted $X$) and one
"backward path" ($Y$)~\cite{FeynmanVernon1963}. 
The action contains two kind of constitutive action parts. 
$S\left[X_i \right]$ and $S\left[Y_i \right]$ are the self-interactions of system $i$ and thus represent the evolution of the density matrix by a single-system Hamiltonian $H_i$. 
$S\left[X_i,X_j \right]$ and $S\left[Y_i,Y_j \right]$ are similarly the interactions between systems $i$ and $j$, neighbours
in the interaction graph, and represent the evolution of the density matrix by a system-system interaction Hamiltonian $H_{ij}$. 

We are interested in the reduced dynamics of the $i$'th system, obtained by tracing out all other system. 
Tracing first out all units except the  $i$'th unit and its neighbours
labelled $j$, we can rewrite Eq. (\ref{eq:evolution}) as 
an evolution equation for the density matrix of system 
$i$
\begin{widetext}
\begin{eqnarray}
\rho_f^{(i)}\left(X_i(t_f),Y_i(t_f)\right) &=&\int D X_i D Y_i \prod_{j \in \partial i  }  D X_j D Y_j \exp \left[ \frac{i}{\hbar} \left(S\left[X_i \right] -S\left[Y_i \right]  +S\left[X_j \right] - S\left[Y_j \right] + S\left[X_i,X_j \right] -S\left[Y_i,Y_j \right]  \right) \right] \nonumber \\
\label{eq:Eq1-v2}
&& \quad \exp \left[ \frac{i}{\hbar} {\cal F}_{\partial i } \left[ \left( X_j,Y_j\right)_{j\in \partial i} \right] \right]  \rho_0^{(i)}\left(x_i(t_i),y_i(t_i)\right) \prod_{j \in \partial i  } \delta (x_j(t_f)-y_j(t_f))\,
\rho_0^{(j)}\left(x_j(t_i),y_j(t_i)\right)
\end{eqnarray}
\end{widetext}
Here and in the following we assume a factorized initial state and trace the final state of all 
states $j\in\partial i$, the set of neighbours of $i$.
The functional 
${\cal F}_{\partial i }$ is the result of integrating over the histories and tracing the final state of
all variables except those in node $i$ and 
 $\partial i$. It is a functional of the histories 
in $\partial i$, but does not know about $i$ itself. One says that variable $i$ has been removed, and its place in the original network has been
replaced by a cavity.

The locally tree-like geometry means that the variables in $\partial i$ are far apart in this new cavity network. They are not independent,
but after $i$ have been removed' their dependence is through many intermediate nodes. The fundamental assumption of the cavity method on the Replica Symmetric (RS)
level is that in a large 
enough network the nodes in $\partial i$ are eventually independent. This means that ${\cal F}_{\partial i }$ 
simplifies as
\begin{equation}
{\cal F}_{\partial i } = \sum_{j\in \partial i} F_{j \rightarrow i } \left[ X_j,Y_j \right] \quad\hbox{(RS cavity assumption)}
\label{eq:cavity-assumption}
\end{equation}
Structurally, \eqref{eq:Eq1-v2} is now a \textit{BP output equation}
where the the $F_{j \rightarrow i }$ play the roles
of \textit{BP messages}.
Such messages, conventionally denoted $m_{j \to i}$ and  $n_{j \to i}$, 
obey recursive equations known as
\textit{BP update equations}.
The BP update equations are illustrated in
Fig.~\ref{fig:BP} (for the definitions and equations,
see Supplementary Information).

\begin{figure*}[h!]
\centering
\includegraphics[scale=0.17]{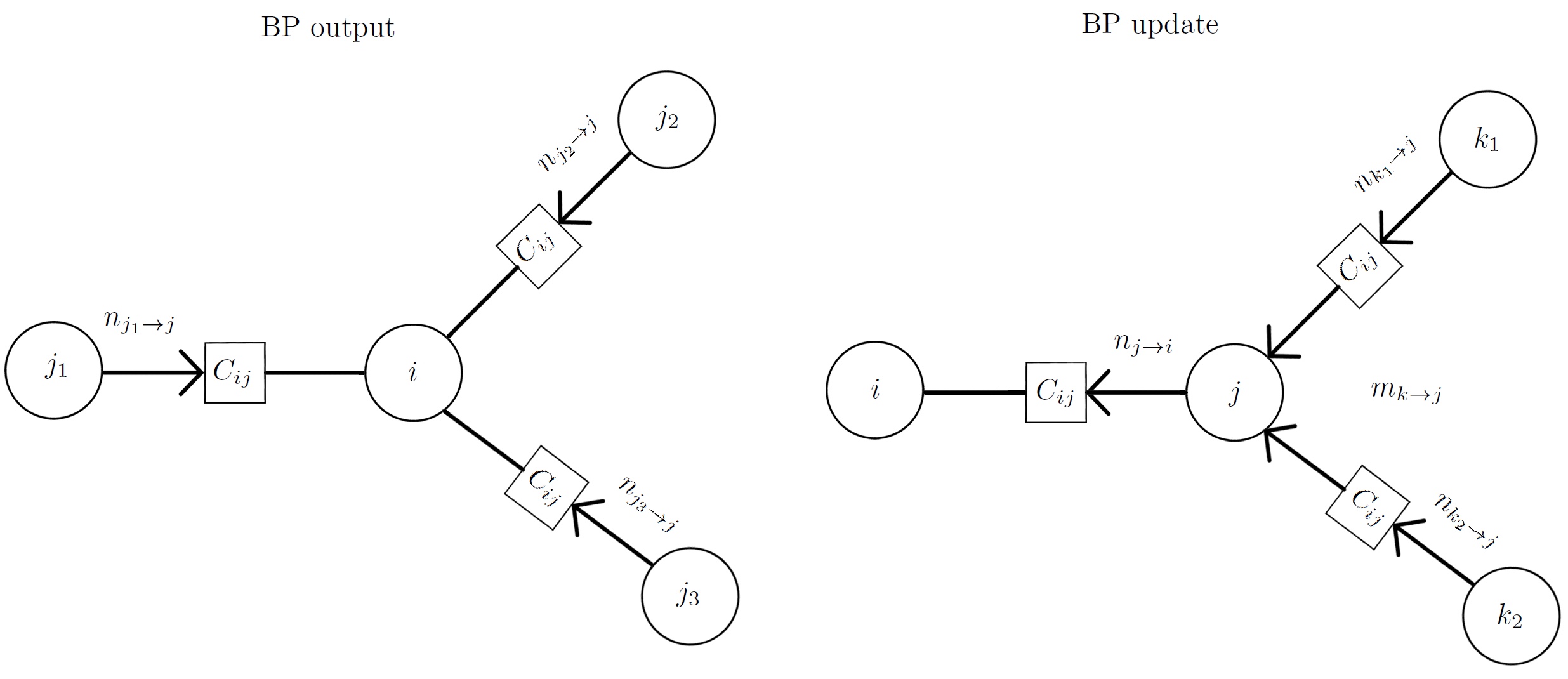}
		\caption{Left panel: illustration of the 
		marginal probability amplitude over 
		local histories as expressed by
		\eqref{eq:Eq1-v2} with cavity assumption
		\eqref{eq:cavity-assumption}.
		Messages $n_{j \to i}$ are equal to 
		$\exp\left(\frac{i}{\hbar}F_{j \rightarrow i } \right)$ where
		$F_{j \rightarrow i }$ appear in
		\eqref{eq:Eq1-v2}, and represent the effect of integrating out nodes
		in the graph subtended from $j$, but not $j$ itself.
		Right panel: illustration of the recursive
		equation satisfied by the messages.
		Nodes labelled $k$ are here coupled to a node labelled $j$, which in turn is 
		coupled to a node labelled $i$. Message $m_{j \to i}$ is obtained 
		by a double path integral over the histories $X_j$ and $Y_j$ of node $j$
		with the bare action $S_j$, the interaction action $S_{ij}$, and the upstream
		influence $F_{j \rightarrow i }$. This message hence represents integrating out all nodes subtended from node $j$, and node $j$ itself.
		Message  $n_{j \to i}$ is on the hand obtained by combining messages $m_{k \to j}$ for $k$ in the neighborhood of $j$ except $i$. 
		In Section \ref{sec:results} we discuss harmonic oscillator networks
		where $n_{j \to i}$ and $m_{j \to i}$ are represented by two pairs
		of kernels, respectively ($k_I^{i\to j},k_R^{i\to j},$) and
		($\tilde{k}_I^{i\to j},\tilde{k}_R^{i\to j},$). The BP update equations
		are then given by
		\eqref{eq:BP-update-for-n-I} and
        \eqref{eq:BP-update-for-n-R} for $m$ to $n$,
        and, under further assumptions, by
        \eqref{eq:transformation-imaginary} and
        \eqref{eq:transformation-real} for $n$ to $m$.
        For further details and the general setting, see Supplementary Information.
}
\label{fig:BP}
\end{figure*}

Both
$m_{j \rightarrow i}$ and $n_{j \rightarrow i}$
are very high-dimensional objects and the update step
is therefore in general quite complex and computationally expensive.
Further assumptions or approximations are needed to
get useful results. A similar problem arises also in
classical dynamic cavity~\cite{Neri2009,Aurell2011,Aurell2012,DelFerraro2015,Barthel2018,Barthel2020,Aurell2017,Aurell2018b,Aurell2019}, though 
there is then only one history per system. 

\section{Uniform harmonic networks}
\label{sec:results}
As a solvable example with interesting and indeed 
unexpected properties, we
now a discuss a uniform random network of
harmonic oscillators which interact linearly.
The action is then 
$\mathcal{S}[X_1, \cdots, X_N]  = 
\int_{t_i}^{t_f} \frac{m}{2} \sum_j \dot{X}_j^{2}-\frac{m \omega^{2}_{0,j}}{2} X^{2}_j+ \frac{1}{2} \sum_{j,i \in \partial j} C_{ij} (X_i- X_j)^2\, dt$
where $m$ is the oscillator mass, $\omega_{0,j}$ is the frequency and
$C_{ij}$ is the spring constant between oscillator $i$ and $j$.
By a change of scale we can take all oscillators of the same
mass.
In this section we will also assume all oscillator 
frequencies the same, and we hence drop the index $j$ on 
frequency $\omega_{0,j}$. 
The single-system action parts are 
 $S\left[X_j \right]=\int_{t_i}^{t_f}  \frac{m}{2} \dot{X}_j^{2}-\frac{m \omega^{2}_0 + \sum_{i \in \partial j} C_{ij} }{2} X^{2}_j  \, dt$, 
 and it is convenient to use the notation $m \omega^2 = m \omega^2_0 + \sum_{i \in \partial j} C_{ij} $.
 The system-system actions are
  $S\left[X_j,X_k \right]=\int_{t_i}^{t_f} \left(-C_{ij}\right) X_jX_k\, dt$. 
Eventually we will in this section consider the case when all $C_{ij}$ are the same.

A system of this type can always be solved by diagonalization. However, as except in one dimension the total Hamiltonian is
partly random from the structure of the locally tree-like graph,
this is not trivial. We will see that the real-time
cavity methods offers a more convenient approach,
and
in fact a way to compute marginals of the diagonalization without
actually performing it.

Influence functionals from an environment of harmonic oscillators are parametrized by two kernels, $k_I$ which describes dissipation, and $k_R$ which describes dispersion~\cite{FeynmanVernon1963}.
We can therefore introduce a pair of kernels 
 $k_I^{i\to j}$ and $k_R^{i\to j}$ to parametrize the cavity influence
 functional $F_{i\to j}$.
It is convenient to introduce an analogous cavity functional 
$\tilde{F}_{i\to j}$ for the $m$-type BP message and its pair
of kernels $\tilde{k}_I^{i\to j}$ and $\tilde{k}_R^{i\to j}$.
The two types of kernels are related by
\begin{eqnarray}
\label{eq:BP-update-for-n-I}
k_I^{j \rightarrow i}(s,t-s) 
&=& \sum_{k\in\partial j\setminus i}
\tilde{k}_I^{k \rightarrow j}(s,t-s) \\
\label{eq:BP-update-for-n-R}
k_R^{j \rightarrow i}(s,t-s) 
&=& \sum_{k\in\partial j\setminus i}
\tilde{k}_R^{k \rightarrow j}(s,t-s)
\end{eqnarray}
which is the $m$ to $n$ part of update scheme illustrated in
Fig.~\ref{fig:BP} for harmonic networks. 

The 1959 PhD thesis of Frank Vernon~\cite{VernonPhD}
contains in Appendix V an analysis of the situation where one oscillator
($i$) interacts with another oscillator ($j$), which
in turn interacts with a bath of oscillators.
The influence of the bath on $j$ is described by
an influence action in the forward and backward paths of
$j$. Integrating out also $j$ then leads to an influence action in the forward and backward paths of $i$.
This \textit{Vernon transform} was recently discussed 
by two of us in \cite{AurellTuziemski2021}.
The same scheme obviously also describes the
$n$ to $m$ part of BP update of our concern here.
An important special case is when the process has been going on for a long time under constant conditions.
In this case the transformation of $k_I$ simplify greatly
on the Laplace transform side, and reads
\begin{eqnarray}
\tilde{k}_I^{j \rightarrow i}(\lambda) &=& \, 
\frac{C_{ij}^2}{2}
G_{0,j}(\lambda)
\left(1- G_{0,j}(\lambda)
{k}_I^{j \rightarrow i}(\lambda)\right)^{-1}
\label{eq:transformation-imaginary}
\end{eqnarray}
where $G_{0,j}(\lambda)=\frac{2}{m}\frac{1}{\lambda^2+\omega_{j}^2}$ 
is twice the response function of a free harmonic oscillator with the parameters
of oscillator $j$.
The imaginary part of the Vernon transform ${\cal V}$
is a nonlinear transformation 
of ${k}_I^{j \rightarrow i}$ to $\tilde{k}_I^{j \rightarrow i}$
which acts on each Laplace transform term separately.
As it does not depend on the real part, and neither does
\eqref{eq:BP-update-for-n-I}, the imaginary kernels
of the cavity influence functionals form a system of updates
closed in themselves. This is also true without the assumptions
that the process has been going on for a long time
under constant conditions.
The real part of the 
Vernon transform ${\cal W}$
is on the other hand a linear transformation 
of ${k}_R^{j \rightarrow i}$ to $\tilde{k}_R^{j \rightarrow i}$
which depends quadratically on $\tilde{k}_I^{j \rightarrow i}$,
\textit{i.e.} on the image of ${\cal V}$.
It simplifies on the Fourier side
\begin{eqnarray}
\tilde{k}_R^{j \rightarrow i}(\nu) &=&
\frac{4}{C_{ij}^2}
|\tilde{k}_I^{j \rightarrow i}(\nu)|^2 k_R^{j \rightarrow i}(\nu),
\label{eq:transformation-real}
\end{eqnarray}
where $\tilde{k}_I^{j \rightarrow i}(\nu)$ can be defined
directly on the Fourier side, or by analytic continuation from
$\tilde{k}_I^{j \rightarrow i}(\lambda)$\footnote{For simplicity we
use the same symbols for the Laplace and Fourier transforms
and time-domain functions.}.
In any case, the real kernels of the cavity influence functions
do not form a system closed in themselves.     
Furthermore, for a process over a finite time ${\cal W}$ also contain
other terms which also depend on $\tilde{k}_I^{j \rightarrow i}$
and on initial conditions (bath temperature),
but not on $k_R^{j \rightarrow i}$, see Supplementary
Information.

For a harmonic locally tree-like network we have thus arrived at
a system of updates of real numbers where
one can look for fixed points.
In the uniform network 
all oscillator frequencies
and all interaction parameters are the same
($\omega_j=\omega$ for all $j$, 
$C_{ij}=C$ for pairs $i$ and $j$), and the size of the neighborhood of each system is the same.
The includes systems on the line, and systems on random regular graphs.
We call the size of the neighborhood of each system
$n$ (the line being $n=2$).
The uniform fixed point where all kernels
everywhere in the network are the same
is then on the Laplace transform side given by
the fixed point of the one-dimensional map
\begin{equation}
\label{eq:iteration-equation}
k_I^{i+1}(\lambda)
= \frac{(n-1)C^2}{2}
{G}_0(\lambda)
\left(1- {G}_0(\lambda)
k_I^{i}(\lambda)\right)^{-1}
\end{equation}
There is always a fixed point of this rational map
for $C$ small enough. When it exists it is given by
\begin{eqnarray}
\label{eq:k-solution-2}
k_I^*(\lambda) 
&=& m\frac{\lambda^2+\omega^2}{4}
\left(1-\sqrt{1-\frac{8(n-1)\, C^2 }{m^2(\lambda^2 +\omega^2)^2}}\right) 
\end{eqnarray}
Given that $\omega^2$ has been defined as 
$\omega^2=\omega_0^2+\frac{nC}{m}$, the 
expression inside the square root in \eqref{eq:k-solution-2}
is a decreasing function of $C$, 
positive
for all $\lambda$ if either $n\geq 8$ or if
$n<8$ and $C$ is less than a critical
value $C^*(n)=\frac{m\omega_0^2}{\sqrt{8(n-1)}-n}$.
For $n<8$  and $C>C^*(n)$ the fixed point still exists for $\lambda$
larger than 
$\lambda^*(C,n)=\omega_0\sqrt{\frac{C}{C^*}-1}$.

The first result on this quite simple example is that
if $C<C^*$
every system in
the uniform network behaves as if interacting with 
the same effective environment.
We call this the ordered phase.
The fixed point kernel $k_I^*$ in the time domain
is illustrated in Fig~\ref{fig:kernel}; a more detailed 
discussion can be found in Supplementary Information.

\begin{figure*}[h!]
\centering
\includegraphics[scale=0.5]{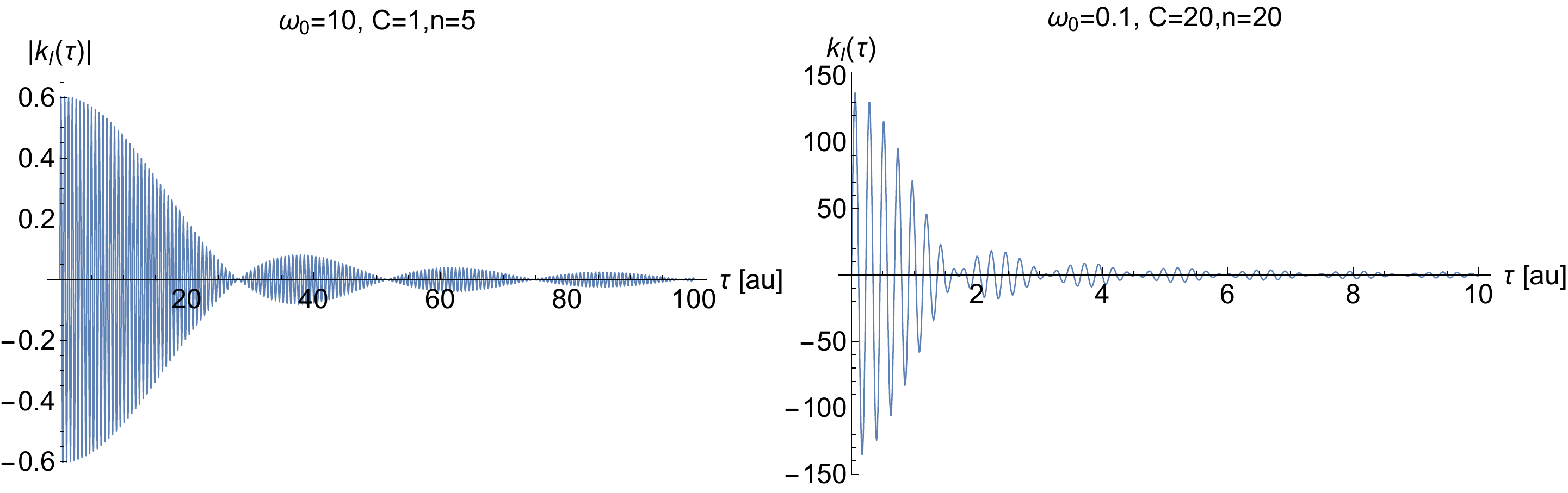}

		\caption{Inverse Laplace transform of the fixed point 
		kernel given by \eqref{eq:k-solution-2}. The parameters are: left panel - $\omega_0=10$, $C=1$, $n=5$, $m=\frac{1}{2}$; right panel -  $\omega_0=0.1$, $C=20$, $n=20$, $m=\frac{1}{2}$.
		One can define an equivalent environment as a set
		of harmonic oscillators giving rise to the same influence functional
		acting on oscillator $j$ as the rest of the harmonic network.
		From the general formula $k_I(\tau)=\sum_b \frac{1}{2m_b\omega_b}\sin\omega_b\tau$
		follows that the spectrum of the equivalent environment, 
		is a band-pass filter with support in the two bands
		$\pm \sqrt{\omega^2 \pm \sqrt{8(n-1)}\frac{C}{m}}$.
		This explains the form of $k_I(\tau)$ which has an oscillatory
		part of frequency $\omega$ and an overlayed breather oscillation.
		Note that a diagonalization in terms of normal modes must also 
		lead to these kind of formulas, but are not trivial to obtain
		even for systems on a line.
}
\label{fig:kernel}
\end{figure*}
For $C>C^*(n)$ and $\lambda<\lambda^*(C,n)$ the BP messages (the
functions $k_I^i(\lambda)$) oscillate
as functions of iteration index $i$. 
In this setting it is 
therefore not consistent to assume that all incoming messages in Fig.~\ref{fig:BP} to be the same; there is nothing to synchronize them.
As we will discuss in the following Section, one can instead assume that each such
$k^{k\rightarrow j}_I(\lambda)$ is a random number drawn from a probability
$P\left(k^{k\rightarrow j}_I(\lambda)\right)$
and check if this distribution is preserved as $P\left(k^{j\rightarrow i}_I(\lambda)\right)$ (Replica Symmetric analysis of the BP update equations).
In this regime, 
every system in
the uniform network then behaves as if interacting with 
an environment, drawn from the same distribution of environments. 
We call this the dynamically disordered phase.
The instances of this phase are quite complex, as
there is no smooth function $f(t)$ which has a Laplace transform
$\tilde{f}(\lambda)$, for which when $\lambda<\lambda^*$
the values are independent random numbers.
The instances $k_I(t-s)$ hence have to be non-smooth mathematical 
distributions.
The seemingly pathological property can be traced back to our neglecting 
the first time in the definition of $k_I(s,t-s)$.
In the time domain $k_I^{i+1}(s,t-s)$ is related to 
$k_I^{i}(s-\tau_I,t-s)$ where $\tau_I$ is a characteristic time of the
response function. When the process starts at time $t_i$
one can only iterate the quantum cavity for 
$i\approx \frac{s-t_i}{\tau_i}$ times before the initial conditions
start to be felt. Therefore, the different components $k_I^i(s,\lambda)$
are actually correlated over a distance in $\lambda$ of size roughly
$\Lambda^{-i}$ where $\Lambda$ is a characteristic expansion rate of 
the Vernon transform, and $k_I^{i}(s,t-s)$ are smooth functions,
albeit quite irregular for $s-t_i$ much greater than $\tau_I$.

When $k_I$ is at the fixed point \eqref{eq:k-solution-2}, the real Vernon transform
\eqref{eq:transformation-real}
is a simple multiplication
\begin{eqnarray}
k_R^{i+1}(\nu) &=& A(\nu)  k_R^{i}(\nu)
\label{eq:transformation-real}
\end{eqnarray}
The multiplier $A(\nu)$ depends on whether the square root in
\eqref{eq:k-solution-2} (for $\lambda=i\nu$) is positive or negative.
The second case pertains to a band of frequencies around 
$\omega$ with width proportional to $C$. The value is
then $2$, while it decreases
down from $2$ away from the band.
In the ordered phase successive iteration of some Fourier components
of $k_R$ hence increase 
without limit.
By the same argument as above this pathological behaviour
can be traced back to neglecting the first time
; actually $k_R(s,\nu)$ only reaches size $2^{\frac{s-t_i}{\tau_i}}$.
Furthermore we have also here neglected additional terms making
${\cal W}$ an affine transformation, see
Supplementary Information.
In the disordered phase the behaviour of ${\cal W}$
is that of random transform; its analysis will be left
to future work.

\section{Disorder and quantum dynamical Replica Symmetry}
\label{sec:RSB}
Understanding disorder in parameters is an important application of the cavity method.
The fixed points of BP will then not be uniform,
but the messages $m_{i\rightarrow j}$ and 
$n_{i\rightarrow j}$ depend on the link in the network.
For harmonic networks this means that the Feynman-Vernon
fixed point kernels $\left(\tilde{k}_I^{i\rightarrow j},\tilde{k}_R^{i\rightarrow j}\right)$ and
$\left(k_I^{i\rightarrow j},k_R^{i\rightarrow j}\right)$ are different for different
pairs $(i,j)$.
We recall how this is analyzed on the level of Replica Symmetry (RS).
An ensemble of such networks and interactions are described by 
probability distributions over the kernel pairs which
obey consistency conditions 
known as RS cavity equations.
As the BP update equation acts on $k_I$ kernel alone it is
convenient to consider separately 
$P(\tilde{k}_I)$ and $Q(k_I)$, which satisfy RS cavity equations
\begin{equation}
Q(k_I) = \sum_k \Lambda_k \int \prod_k \big[ d \hat{k}_I^k P( \hat{k}_I^k) \big] \delta( k_I - \sum_k \hat{k}_I^k)
\label{eq:RS-Q-def}
\end{equation}
and
\begin{equation}
P(\hat{k}_I) = E_{C}  \int d \hat{k}_I Q(k_I) \delta( \hat{k}_I - {\cal V}(k_I) )
\label{eq:RS-P-def}
\end{equation}
where $\Lambda_k$ is the probability of the neighbourhood to be of size $k$,
$E_{c}$ represents the average over the coupling distribution and ${\cal V}$ is the $n$ to $m$ step of the
BP update determined by the Vernon transform.
The kind of solution discussed above  is in this formulation described by
$P$'s and $Q$'s that are delta functions.

As we also know from the previous section,
even with no disorder in the parameters there is
a dynamically disordered phase of the quantum cavity in a uniform
network.
There is then only one term in the sum in 
\eqref{eq:RS-Q-def}, and no average over
parameters in
\eqref{eq:RS-P-def}.
Nevertheless, if the Vernon transform does not have a fixed
point, the effect is similar. 
We note that in this setting \eqref{eq:RS-Q-def} 
and \eqref{eq:RS-P-def} can be combined on the Fourier side 
($\hat{P}(l) = \int e^{ilx} P(x) dx$) to one
update equation
\begin{equation}
\hat{P}^{i+1}(l) = \int dl' F(l,l') \left(\hat{P}^{i}(l')\right)^{n-1}
\label{eq:RS-P-def-2}
\end{equation}
with the transfer function
\begin{equation}
F(l,l') = \frac{1}{2\pi}\int e^{i\left(l'{\cal V}(x)+lx\right)}\, dx
\label{eq:RS-P-def-2}
\end{equation}
For systems on a line ($n=2$) this iteration is simply the 
Perron-Frobenius transform corresponding to the dynamics given by
$x\to {\cal V}(x)$.

\section{Discussion and outlook}
\label{sec:discussion-outlook}
We have in this work introduced a new real-time version of the 
quantum cavity method. We have shown that when all systems in a network 
are harmonic oscillators interacting linearly, 
this real-time quantum cavity can be represented 
as transforms of Feynman-Vernon kernels. We have also shown that 
for uniform harmonic networks
where all interaction coefficients in the network are the same, 
there is an ordered phase where
can solve explicitly for a fixed point of transformations.
A single quantum harmonic oscillator interacting
with a network of quantum harmonic
oscillators in the topology of a  $n$-regular random graph then 
behaves as if under the influence of a dissipation kernel of
finite spectral support, and eventually arbitrarily strong decoherence,
if the dynamics has been going on for a long time.
Such a network will hence eventually behave entirely classically.

We have also discussed disorder and Replica Symmetry for real-time
quantum cavity. 
We have pointed out that also the uniform harmonic network has
a dynamically disordered phase, in the absence of any disorder in the 
parameters.
Disorder in model parameters induce an Anderson transition
on locally tree-like graphs \cite{Abou_Chacra1973,MillerDerrida1994,Parisi2019}.
For small disorder states are delocalized and influence propagate though the
network, while for large disorder states are localized. 
We conjecture that such a transition also takes place
for our real-time problem also.
Each system would then perceive the network as an instance of
an ensemble of effective environments. To study the properties
of these effective environments is an important task for the future.

Finally, the examples we have considered are
networks of harmonic oscillators,
as they yield explicit solutions of the real-time quantum cavity in
closed form.
While the challenges in extending
these investigations to qubits or
other systems are significant there are ways ahead. The first
is that the Feynman-Vernon 
transform 
is defined for environments that
are not harmonic oscillators; the 
difference being that the Feynman-Vernon
action will then have terms cubic, quartic
etc in the system variables. The kernels
of those higher terms are cumulants
of environment correlation functions
which vanish for harmonic oscillator baths \cite{AurellKawaiGoyal2020}.
While keeping an an infinite tower of
higher-order Feynman-Vernon kernels
will surely be impractical, one could
consider truncations, such as that
the action at every step remains quadratic.
Analogous truncations have found
many applications in classical 
information processing 
\cite{Sudderth2003,Bickson2009}.
Another direction is that
path integrals for spins
have been developed,
and it is conceivable that
transforms analogous to the Vernon
transform could be developed for
them also.

\section*{Acknowledgments}
  EA thanks Foundation for Polish Science through TEAM-NET project (contract no. POIR.04.04.00-00-17C1/18-00) 
  for support during the initial phase of this work, 
  Swedish Research Council grant 2020-04980 during its 
  completion phase.
  JT was supported by the European Research
Council grant 742104.

\begin{widetext}

\section*{Supplementary Information}
\tableofcontents

\section{Details the real-time quantum cavity method for harmonic oscillator networks.}
\label{sec:SI-Vernon}
In this section we consider a locally tree-like graph 
where in each vertex resides a quantum harmonic oscillator.
These oscillators interact linearly as indicated by the graph structure.

The starting point of real-time quantum cavity method are the Feynman-Vernon functionals
from integrating out a whole tree subtended by one node.
As discussed in the main paper these are of two types,
conventionally in the cavity literature called "$n$-type" and and "$m$-type messages.
We introduce parametrizations so that for a harmonic
network they read

\begin{eqnarray}
n_{j \rightarrow i} [X_j,Y_j]= e^{\frac{i}{\hbar} F_{j \rightarrow i} [X_j, Y_j ]} &=&
\exp \left\{\frac{i}{\hbar} \int^{T}_{\tau} \int^{t}_{\tau}
k_I^{j \rightarrow i}(s,t-s) \left(X_j(t)-Y_j(t)\right) \left(X_j(s)+Y_j(s)\right)   d t d s \right\} \nonumber \\ 
&&\exp \left\{- \frac{1}{\hbar} \int^{T}_{\tau} \int^{t}_{\tau}
k_R^{j \rightarrow i}(t,s) \left(X_j(t)-Y_j(t)\right) \left(X_j(s)-Y_j(s)\right) \right\} \\
m_{j \rightarrow i} [X_i,Y_i] = e^{\frac{i}{\hbar} \tilde{F}_{j \rightarrow i} [X_i, Y_i ]} &=&
\exp \left\{\frac{i}{\hbar} \int^{T}_{\tau} \int^{t}_{\tau}
\tilde{k}_I^{j \rightarrow i}(s,t-s) \left(X_i(t)-Y_i(t)\right) \left(X_i(s)+Y_i(s)\right)   d t d s \right\} \nonumber \\ 
&&\exp \left\{- \frac{1}{\hbar} \int^{T}_{\tau} \int^{t}_{\tau}
\tilde{k}_R^{j \rightarrow i}(t,s) \left(X_i(t)-Y_i(t)\right) \left(X_i(s)-Y_i(s)\right) \right\} 
\end{eqnarray}
The kernels $k^{j\to i}$ (symbol without tilde)
in $n_{j \rightarrow i}$ 
multiply histories pertaining to 
the ingress node (node $j$). They represent the effect
of integrating out the histories of the systems in all
nodes
neighbours to $j$ or subtended from neighbours of $j$,
except node $i$ and nodes subtended from $i$.
The kernels $\tilde{k}^{j\to i}$ (symbol with tilde)
in $m_{j \rightarrow i}$ 
multiply histories pertaining to 
the egress node (node $i$). They represent the effect
of integrating out the histories of the systems in
node $j$ and nodes subtended from $j$.

One relation between $n$-messages and $m$-messages follow
from one of the most basic properties of influence 
functionals; that influence functions from
disjoint environments multiply. In our case
we write this as
\begin{equation}
\label{eq:BP-update-n}
    n_{j \rightarrow i}\left[ X_j,Y_j \right] \equiv  \exp \left[ \frac{i}{\hbar} F_{j \rightarrow i }\left[ X_j,Y_j \right] \right] = \prod_{k \in \partial j \backslash i  } m_{k \rightarrow j}\left[ X_j,Y_j \right] \equiv \exp \left[ \frac{i}{\hbar} \sum_{k \in \partial j \backslash i  } \tilde{F}_{k \rightarrow j }\left[ X_j,Y_j \right] \right] 
\end{equation}
which for the kernels translate to
\begin{eqnarray}
\label{eq:BP-update-for-n-I}
k_I^{j \rightarrow i}(s,t-s) 
&=& \sum_{k\in\partial j\setminus i}
\tilde{k}_I^{k \rightarrow j}(s,t-s) \\
\label{eq:BP-update-for-n-R}
k_R^{j \rightarrow i}(s,t-s) 
&=& \sum_{k\in\partial j\setminus i}
\tilde{k}_R^{k \rightarrow j}(s,t-s)
\end{eqnarray}
The other relation between $n$-messages and $m$-messages 
follow from integrating out the histories of the system in node $j$
and reads in general
\begin{eqnarray}
\label{eq:BP-update-m}
  m_{k \rightarrow j}\left[ X_j,Y_j \right] &\equiv&  \int D X_k D Y_k \exp \left[ \frac{i}{\hbar} \left(S\left[X_k \right] -S\left[Y_k \right] +  S\left[X_j,X_k \right] -S\left[Y_j,Y_k \right] \right) \right] \nonumber \\
  && \qquad n_{k \rightarrow j} \left[ X_k, Y_k \right]
  \delta (x_k(t_f)-y_k(t_f))\,  \rho_0^{(k)}\left(x_k(t_i),y_k(t_i)\right)
\end{eqnarray}
For one harmonic oscillator degree of freedom in node $j$ and with interactions as considered here,
this is more explicitly
\begin{eqnarray}
e^{\frac{i}{\hbar} \tilde{F}_{j \rightarrow i} [X_i, Y_i ]} &= &\int \exp \left\{\frac{i}{\hbar} \int^{T}_{\tau}\left[\frac{m}{2}\left(\dot{X}_j^{2}-\dot{Y}_j^ 2\right)-\frac{m \omega^{2}}{2}\left(X^{2}_j-Y^{2}_j\right)+C_{ij} X_i X_j -C_{ij}Y_i Y_j\right] d t \right\} 
e^{\frac{i}{\hbar} F_{j \rightarrow i} [X_j, Y_j ]} \nonumber\\
&&
\delta\left(X_j(T)-Y_{j}(T)\right)  
\frac{1}{N} \exp \left[ -\left( A x^2 + 2B x y +C y^2 \right)/2 \right]
{\cal D}  X_j {\cal D} Y_j\, dX_j(\tau) dY_j(\tau) \, d X_j(T)\, d Y_j(T) 
\label{eq:Vernon-transform-iteration}
\end{eqnarray}
Lowercase letters ($x$ and $y$) 
in above stand for the initial data  on $j$, \textit{i.e.}  $X_j(\tau)$ and $Y_j(\tau)$,
and uppercase letters ($X$ and $Y$) 
will from now on stand for the final data $X_j(T)$ and $Y_j(T)$. 
The core problem is to translate \eqref{eq:Vernon-transform-iteration} into a transformation of kernels.
The geometry of passing of message of types $m$ and $n$ is  illustrated by Fig.~\ref{fig:BP} 

We have required that the initial state in \eqref{eq:Vernon-transform-iteration} is Gaussian
normalized by
\begin{equation}
    \hbox{Tr}\left[\rho^{(j)} (\tau)\right] = \frac{1}{N} \int dx \exp \left[ -\left( A  + 2B +C \right)x^2/2 \right] = 1
\end{equation}
and we will later require that it is symmetric, $A=C$.
Eq.~\eqref{eq:Vernon-transform-iteration}
is then precisely the kind of 
iterated path integral studied by Vernon
in Appendix V of his PhD thesis~\cite{VernonPhD}
and which we call the
\textit{Vernon transform}.
We can therefore immediately 
write down
\begin{eqnarray}
\tilde{k}_I^{j \rightarrow i} &=& \, {\cal V}\left[k_I^{j\to i}, C_{ij}\right]  \label{eq:transformation-imaginary} \nonumber \\
&=& 
C_{ij}(t)\, C_{ij}(s)\, G^{j\to i}(t,s-t)  \label{eq:transformation-imaginary-element}
\end{eqnarray}
\noindent where the response function
and the friction kernel  satisfies the twinning relation
\begin{eqnarray}
\label{eq:G-twinning}
G^{j\to i}(t,s-t) &=&  G_j^{(0)}(s-t)+ \int_{t}^{T} dt_1 \int_{t_1}^{T} dt_2\, G_j^{(0)}(t_1-t)\nonumber \\
&& \quad k_{I}^{k \rightarrow j}(t_1,t_2-t_1)\, G^{j\to i}(t_2,s-t_2) 
\end{eqnarray}
In above $G_j^{(0)}(s-t)$ is \textbf{twice} the response function of oscillator $j$ without 
the friction kernel given by $k_I^{j\to i}$. For derivations, see
Sections~\ref{app:Vernon-time-and-Laplace}
and~\ref{app:G}.

The real side of the Vernon transform  ${\cal W}$ is in general the sum of  three terms. All three depend quadratically on the response function $G^{j\to i}$, one depends linearly on  $k_R^{j \rightarrow i}$ and two terms do not depend on $k_R^{j \rightarrow i}$.

With $A'=\frac{1}{2}\left(A+B\right)$
and $C'=\frac{1}{2}\left(A-B\right)$
characterizing the initial symmetric state of
oscillator $j$ the three terms read

\begin{eqnarray}
\tilde{k}_R^{j \rightarrow i}(t,s) &=& 
C_{ij}(t)\, C_{ij}(s)\, 
\int_{\tau}^{t}\int_{\tau}^{s}
k_R^{j \rightarrow i}(t',s')\, G^{j\to i}(t',t-t')\, G^{j\to i}(s',s-s')\,  ds' dt' \, +\,
\nonumber \\
&&  C_{ij}(t)\, C_{ij}(s)\,
\left(\hbar C'  G^{j\to i}(\tau,t-\tau)\, 
G^{j\to i}(\tau,s-\tau) 
 +\,  
\frac{1}{\hbar A'} 
\frac{dG^{j\to i}(r,t-r)}{dr}|_{r=\tau}\,\, 
\frac{dG^{j\to i}(r,s-r)}{dr}|_{r=\tau}\right) 
\label{eq:transformation-real-full}
\end{eqnarray}

The last two terms 
in \eqref{eq:transformation-real-full}
stem from the initial condition
of oscillator $j$.
They are in fact the same terms that give rise
to the quantum noise kernel $k_R$ in
Vernon's original derivation of the
Feynman-Vernon kernels (\cite{VernonPhD}, Appendix I).
Under conditions discussed in
Section~\ref{app:Vernon-time-and-Laplace}
these two terms vanish in several settings
when the process goes on for an 
infinite time ($\tau=-\infty$).
Also, when $G^{j\to i}$ behaves as
the response function of a damped
harmonic oscillator it has finite 
memory, and the two last  terms  in \eqref{eq:transformation-real-full} are boundary contributions  which only matter in the beginning of the
process (both times $t$ and $s$ close to $\tau$).

Furthermore, the Vernon transforms ${\cal V}$ and ${\cal W}$ simplify considerably
on the Laplace/Fourier side when the interaction coefficients $C_{\alpha}$ (all pairs $\alpha$) do not depend on time, all real kernels $k_R(\cdot,t)$ only depend 
on the second time argument, and all imaginary kernels $k_R(t,s)$ only depend on the time difference.

The twinning equation can thus be written on the Laplace transform side  as
\begin{eqnarray}
\label{eq:G-twinning-Laplace}
\tilde{G}^{j\to i}(\lambda) &=&  \tilde{G}_j^{(0)}(\lambda)
\left(1+ \tilde{k}_{I}^{k \rightarrow j}(\lambda)\, \tilde{G}^{j\to i}(\lambda)\right) 
\end{eqnarray}
The above form was used for the
fixed point calculations in the main body of the paper.

\section{The Vernon transform in the time domain and on the Laplace transform side}
\label{app:Vernon-time-and-Laplace}
This Section contains further details on the
calculation outlined in the previous section.
It starts with a presentation
of the Vernon transform
(\cite{VernonPhD}, Appendix V); this
material can also be found in 
\cite{AurellTuziemski2021}.
In contrast to that earlier presentation
we keep the time dependence throughout,
to eventually write the Vernon transform
for a time-stationary situation
on the Laplace side.

The point of departure is 
\eqref{eq:Vernon-transform-iteration},
the 
path integral expressing the Vernon
transform.
For illustration, for the moment we do not require the initial state to be symmetric.
One introduces the new variables
\begin{eqnarray}
&&\bar{X}_j(t) = X_j(t)+Y_j(t) \\ 
&&\Delta_j(t) = X_j(t)-Y_j(t)
\end{eqnarray}
and similarly for the target node $i$ and the initial and final state on $j$. 
In this way one can write

\begin{eqnarray}
e^{\frac{i}{\hbar}\tilde{F}_{j \rightarrow i} [\bar{X}_i, \Delta_i ]} = && \int \delta\left(\Delta_j^T\right) \exp \left\{\frac{i}{\hbar} \int^{T}_{\tau}\left[\left(\frac{m}{2}\dot{\bar{X}}_j \dot{\Delta}_j  -\frac{m \omega^{2}}{2}\bar{X}_j \Delta_j+\frac{C_{ij}}{2} \bar{X}_i \Delta_j + \frac{C_{ij}}{2} \Delta_i \bar{X}_j   \right)   \right] d t \right\} \nonumber \\ 
&& \qquad \exp \left\{\frac{i}{\hbar} \int^{T}_{\tau} \int^{t}_{\tau} k_I^{j \rightarrow i}(s,t-s) \Delta_j(t)\bar{X}_j(s)  d t d s - \frac{1}{\hbar} \int^{T}_{\tau} \int^{t}_{\tau}  k_R^{j \rightarrow i}(t,s) \Delta_j(t) \Delta_j(s) dt ds \right\} \nonumber \\
&& 
\frac{1}{N} \exp  \left[ -\left(A' (\bar{X}_j^\tau)^2 + 2B' \bar{X}_j^\tau \Delta_j^\tau + C' (\Delta_j^\tau)^2\right)/2 \right]\, 
     {\cal D} \bar{X}_j {\cal D} \Delta_j\, d \Delta_j^{\tau} d \Delta_j^{T} d \bar{X}_j^\tau d \bar{X}_j^T
   \end{eqnarray}
   where 
$A'=\frac{1}{4}\left(A+2B+C\right)$
$C'=\frac{1}{4}\left(A-2B+C\right)$, and  $B'=\frac{1}{2}\left(A-C\right)$.
Requiring that the initial state does not mix $\bar{x}_j$ and $\Delta_j$ leads to
$B'=0$ and the expressions used in \eqref{eq:transformation-real-full} above. 
Those are $A'=\frac{1}{2}\left(A+B\right)$ and
$C'=\frac{1}{2}\left(A-B\right)$.
Note that the normalization  
$\frac{1}{N}\int e^{- \frac{1}{2}(A' \bar{x}^2)}d\bar{x} = 2$.
A thermal state at inverse temperature
$\beta$ has $A=C=\frac{m_j\omega_j}{\hbar}\coth{\beta\hbar\omega_j}$
and $B=-\frac{m_j\omega_j}{\hbar}\sinh^{-1}{\beta\hbar\omega_j}$,
and hence $A'=\frac{m_j\omega_j}{2\hbar}\tanh{\frac{\beta\hbar\omega_j}{2}}$ and $C'=\frac{m_j\omega_j}{2\hbar}\coth{\frac{\beta\hbar\omega_j}{2}}$. 
The assumption of an initially symmetric Gaussian state is hence equivalent to assuming an initial thermal
state where 
the two parameters $A'$ and $C'$
set a length scale 
$\ell=\frac{2\hbar}{m_j\omega_j}=\frac{1}{\sqrt{A'C'}}$
and an inverse temperature
$\beta=\frac{2}{\hbar\omega_j}\tanh^{-1}{\sqrt{\frac{A'}{C'}}}$.
We will drop the primes on $A'$ and $C'$
from now on. 

The first step is now to integrate the term $\dot{\bar{X}}_j\dot{\Delta}_j$ by parts which gives 

 \begin{eqnarray}
  \int_{\tau}^{T} \dot{\bar{X}}_j \dot{\Delta}_j dt &=& 
\dot{\Delta}_j^T \bar{X}_j^T - \dot{\Delta}_j^\tau \bar{X}_j^\tau - \int_{\tau}^{T} \bar{X}_j \ddot{\Delta}_j dt
\end{eqnarray}
from which follows
\begin{eqnarray}
  e^{\frac{i}{\hbar}\tilde{F}_{j \rightarrow i} [\bar{X}_i, \Delta_i ]} = &&\frac{1}{2}\int \delta\left(\Delta_j^T\right)
\exp \left\{\frac{i}{\hbar} \int^{T}_{\tau}\left[
  \left(-\frac{m}{2}\bar{X}_j \ddot{\Delta}_j  -\frac{m \omega^{2}}{2}\bar{X}_j \Delta_j+\frac{C_{ij}}{2} \bar{X}_i \Delta_j + \frac{C_{ij}}{2} \Delta_i \bar{X}_j   \right)
  \right] d t \right\} \nonumber \\
&& \qquad \exp \left\{\frac{i}{\hbar} \int^{T}_{\tau} \int^{t}_{\tau} k_I^{j \rightarrow i}(s,t-s) \Delta_j(t)\bar{X}_j(s)  d t d s - \frac{1}{\hbar} \int^{T}_{\tau} \int^{t}_{\tau}  k_R^{j \rightarrow i}(t,s) \Delta_j(t) \Delta_j(s) dt ds \right\} 
\nonumber \\&& 
\frac{1}{N} \exp  \left[ -\left(A' \bar{X_j^\tau}^2  + C' (\Delta_j^\tau)^2\right)/2  \right]\, \exp \frac{i m}{2 \hbar} ( \dot{\Delta}_j^T \bar{X}_j^T - \dot{\Delta}_j^\tau  \bar{X}_j^\tau )
     {\cal D} \bar{X}_j {\cal D} \Delta_j\, d \Delta_j^{\tau} d \Delta_j^{T} d \bar{X}_j^\tau d \bar{X}_j^T
\label{eq:App-A-point-of-departure}
\end{eqnarray}
The pre-factor $\frac{1}{2}$ is the Jacobian 
of the change of variables at the initial 
time; the other (functional) Jacobian is
included in the path integral measure.
The initial state is 
Gaussian and we can integrate over 
$\bar{X}_j$.
This gives
$\frac{1}{N}\int d \bar{X}_j^\tau e^{-\frac{1}{2} A' \bar{X_j^\tau}^2 - \frac{i m}{2 \hbar}\dot{\Delta}_j^\tau  \bar{X}_j^\tau } = 2e^{-\frac{m^2}{8\hbar^2 A'} (\dot{\Delta}_j^\tau)^2}$
where the factor $2$ cancels in above.
The corresponding integral over the
final state fixes the final velocity for $\Delta_j$, that is $\int d \bar{X}_j^T \exp^{ \frac{i m}{2 \hbar}\dot{\Delta}_j^T \bar{X}_j^T } = \delta (\dot{\Delta}_j^T)$. 
The remaining integrals
over $\bar{X}_j$
at intermediate times give a 
delta-functional

\begin{equation}
\delta(g(\Delta_j)(t))
\end{equation}
where

\begin{eqnarray}
\label{eq:g-def}
g(\Delta_j)(t) &=& \frac{m}{2} \ddot{\Delta}_j(t)  +\frac{m \omega^{2}}{2}\Delta_j(t) -  \frac{C_{ij}}{2} \Delta_i(t) - \int^{T}_{t} k_I^{j \rightarrow i}(t,s-t) \Delta_j(s) d s.
\end{eqnarray}
The integral over the deviation path 
$\Delta_j(t)$ hence has support on a classical path which 
satisfies 
final conditions $\Delta_j = \dot{\Delta}_j= 0$, and equations of motion $g(\Delta X_j)(t)=0$.
It is convenient to call this auxiliary path $Q(t)$. 
The double path integral in
\eqref{eq:App-A-point-of-departure}
hence gives
\begin{eqnarray}
\frac{i}{\hbar} \tilde{F}_{j \rightarrow i} [\bar{X}_i, \Delta X_i ] &= &
\frac{i}{\hbar} \int_{\tau}^{T}
\frac{C_{ij}(t)}{2}\bar{X}_i(t) Q(t) dt -\frac{1}{2\hbar} \int_{\tau}^{T}\int_{\tau}^{T}
k_R^{j \rightarrow i}(t,s) Q(t)Q(s)  -\frac{1}{2}C q^2  -\frac{m^2}{8\hbar^2 A} (\dot{q})^2  
\label{eq:transformation-def1}
\end{eqnarray}
where by $q$ we mean the initial position of the auxiliary classical path, \textit{i.e.} $Q(\tau)$.
$Q(t)$ depends on 
the deviation path $\Delta X_i(s)$ for all values of $s$ larger
than $t$. This is because $Q$ satisfies final conditions
as $s=T$
while its initial conditions at $s=\tau$
are not given.
It is further clear that 
$Q(t)$ is a linear 
functional $\Delta X_i(s)$ for $s\in \left[t,T\right]$.
This linear functional can be represented by a kernel
\begin{equation}
\label{eq:G-def}
Q(t) = \int_t^{T}G(t,s-t)\,  C_{ij}(s) \Delta X_j(s)\, ds
\end{equation}
where
\begin{equation}
G(t,s-t) =  G_0(s-t)+ \int_{t}^{T} dt_1 \int_{t_1}^{\infty} dt_2 G_0(t_1-t) k_{I}^{k \rightarrow j}(t_1,t_2-t_1) G(t_2,s-t_2) 
\end{equation}
As shown in Section~\ref{app:G} there is at least as a formal power series
there is a kernel which satisfies $G(t,s-t)=0$ for $s<t$.
Substituting \eqref{eq:G-def}
in \eqref{eq:transformation-def1}
we have the kernels of the transformed Feynman-Vernon
action as
\begin{eqnarray}
\tilde{k}_I^{j \rightarrow i}(t,s-t) &=& \frac{1}{2}\, C_{ij}(t)\, C_{ij}(s)\, G(t,s-t)  
\label{eq:transformation-imaginary}
\\
\tilde{k}_R^{j \rightarrow i}(t,s) &=& 
C_{ij}(t)\, C_{ij}(s)\, 
\int_{\tau}^{t}\int_{\tau}^{s}
k_R^{j \rightarrow i}(t',s')\, G(t',t-t')\, G(s',s-s')\,  ds' dt' 
\nonumber \\
&& \quad +\, C_{ij}(t)\, C_{ij}(s)\, \hbar C 
G(\tau,t+\tau)\,
G(\tau,s+\tau)
\nonumber \\
&& \qquad +\,  C_{ij}(t)\, C_{ij}(s)\, 
\frac{1}{\hbar A} 
\frac{dG(r,t-r)}{dr}|_{r=\tau}\,\, 
\frac{dG(r,s-r)}{dr}|_{r=\tau}\,\, 
\label{eq:transformation-real}
\end{eqnarray}
The last two terms are zero if the auxiliary path
$Q(t)$ returns to rest at the origin at $t\to\tau$. 
This should be so whenever the kernel
$k_I^{j \rightarrow i}(t,s-t)$ behaves as friction and when the process
goes on for infinite time ($\tau=-\infty$),
and when the drive
($C_{ij}(t)$) vanishes before some
turn-on time $t_i$.  
In any case, if the response 
function $G(\tau,t+\tau)$
has essentially finite support in
$t$, these two terms will only 
give a boundary contribution,
and will not matter when $t$ and $s$
are sufficiently larger than $\tau$.

In this way we have formally established the BP update equation as an integral transformation  on Feynman-Vernon kernels given by \eqref{eq:transformation-imaginary} and the first line of \eqref{eq:transformation-real}. Equations, (\ref{eq:transformation-imaginary}) and (\ref{eq:transformation-real}), can be closed through the use of (\ref{eq:BP-update-n}). 
Concretely we write them as

\begin{eqnarray}
  k_I^{j \rightarrow i}(t,s-t)& =& \sum_{k \in \partial j /i}\tilde{k}_I^{k \rightarrow j}(t,s-t) = \sum_{k \in \partial j /i} \frac{1}{2}\, C_{kj}(t)\, C_{kj}(s)\, G(t,s-t)  
  \nonumber \\
  k_R^{j \rightarrow i}(t,s)& =& \sum_{k \in \partial j /i }\tilde{k}_R^{k \rightarrow j}(t,s) = \sum_{k \in \partial j /i} C_{kj}(t)\, C_{kj}(s)\, 
\int_{-\infty}^{t}\int_{-\infty}^{s}
k_R^{k \rightarrow j}(t',s')\, G(t',t-t')\, G(s',s-s')\,  ds' dt'
\label{eq:messpassings}
\end{eqnarray}
Note that the information about the graph structure is given through the introduction of the $\sum_{k \in \partial j /i }$. 

\section{Representation of the kernel $G$}
\label{app:G}
The goal of this Section is to derive
an explicit representation of the kernel
$G$ defined in \eqref{eq:G-def} in the preceding Section.
This kernel is to relate a function $Q(t)$ 
to a source term $C_{ij} (s) \Delta X_i (s)$
for all $s>t$. Values of the source 
at times $s$ earlier than $t$ have no influence
on $Q(t)$.
The solution must therefore satisfy 
$G(t,s-t)=0$ when $s<t$.

$Q(t)$ is determined by the equation of motion of an externally driven harmonic oscillator with non-Markovian damping (given in preceding appendix as \eqref{eq:g-def}
(and below as \eqref{eq:oscwithkernel})
and final conditions $Q(T) = \dot{Q}(T)= 0$.
For an infinite time interval
($\tau=-\infty$ and $T=\infty$) 
it is convenient to assume that
$C_{ij} (s)$ vanishes for $s>t_f$ as well as for $s<t_i$. The first means that $Q(t)$
must also vanish for $t>t_f$. 
The second means that 
for those values 
$Q$ satisfies an autonomous integro-differential
equation without drive. That is, if we know
$Q$ in the interval $[t_i:t_f]$ then
$Q(t)$ for $t<t_i$ follows as a consequence.

For convenience we restate
the equation 
satisfied by $Q$
representing oscillator $k$
driven by oscillator $j$:

\begin{equation}
  \frac{m}{2} \ddot{Q} + \frac{m \omega^2}{2} Q -\frac{1}{2} C_{kj}(t) \Delta_j(t) - \int_t^\infty ds k_I^{k \rightarrow j}(t,s-t) Q(s) = 0
  \label{eq:oscwithkernel}
\end{equation}
To emphasize that 
$k_I^{k \rightarrow j}(t,s-t)$
we write out explicitly
a Heaviside function
$\Theta(s-t)$.
We start introducing the Fourier transform of

\begin{equation}
  Q(t) = \frac{1}{2 \pi} \int_{-\infty}^\infty d \nu e^{-i \nu t} \hat{Q}(\nu) 
\end{equation}
and

\begin{equation}
\label{eqi:khat}
  k_I^{k \rightarrow j}(t,s-t) \Theta[s-t] =  \frac{1}{2 \pi} \int_{-\infty}^\infty d \mu e^{-i \mu (s-t)} \hat{k}_{I^{-}}^{k \rightarrow j}(t,\mu)
\end{equation}
where $\mu$ should have an infinitesimal positive imaginary part. Substituted in Eq. \eqref{eq:oscwithkernel} this leads to
\begin{equation}
\frac{1}{2 \pi}  \int d\nu e^{-i \nu t} \Bigg[ \frac{m}{2}(-\nu^2 + \omega^2) \hat{\Delta}_k(\nu) -\frac{1}{2} \Big[C_{kj} \Delta_j\Big]_\nu - \hat{k}_{I^{-}}^{k \rightarrow j}(t,-\nu) \hat{\Delta}_k(\nu) \Bigg]= 0
  \label{eq:oscwithkernelkspace}
\end{equation}
where $\nu$ should have an infinitesimal negative imaginary part.

Integrating over time, $\int dt e^{i \kappa t}$ 

\begin{equation}
\frac{1}{2 \pi}  \int dt e^{i \kappa t} \int d\nu e^{-i \nu t} \Bigg[ \frac{m}{2}(-\nu^2 + \omega^2) \hat{\Delta}_k(\nu) -\frac{1}{2} \Big[C_{kj} \Delta_j\Big]_\nu - \hat{k}_{I^{-}}^{k \rightarrow j}(t,-\nu) \hat{\Delta}_k(\nu) \Bigg]= 0
\end{equation}
and defining

\begin{equation}
\label{eq:kidhat}
\hat{\hat{k}}_{I^{-}}^{k \rightarrow j}(\kappa-\nu,-\nu) = \int_{-\infty}^{\infty} dt e^{i (\kappa -\nu )t} \hat{k}_{I^{-}}^{k \rightarrow j}(t,-\nu) 
\end{equation}
one finds:

\begin{equation}
  \Bigg[ \frac{m}{2}(-\kappa^2 + \omega^2) \hat{Q}(\kappa) -\frac{1}{2} \Big[C_{kj} \Delta_j\Big]_\kappa - \frac{1}{2 \pi} \int_{-\infty}^{\infty} d\nu \hat{\hat{k}}_{I^{-}}^{k \rightarrow j}(\kappa-\nu,-\nu) \hat{Q}(\nu) \Bigg] = 0
\end{equation}
where $\kappa$ should have an infinitesimal negative imaginary part.

We can re-write the preceding expression as
\begin{equation}
  \hat{Q}(\kappa) = \frac{\frac{1}{2} \Big[C_{kj} \Delta_j\Big]_\kappa} { 
    \frac{m}{2}( \omega^2 - \kappa^2 )} +
  \frac{ \frac{1}{2 \pi} \int_{-\infty}^{\infty} d\nu \hat{\hat{k}}_{I^{-}}^{k \rightarrow j}(\kappa-\nu,-\nu) \hat{Q}(\nu)}  {\frac{m}{2}(\omega^2 - \kappa^2 )}
\end{equation}
which is a Fredholm singular integral. This can be written as
\begin{equation}
  \hat{Q}(\kappa) = f(\kappa) + \lambda \int_{-\infty}^{\infty} 
  d\nu K(\kappa, \nu) \hat{Q}(\nu)
\end{equation}
Applying the method of successive
iterated approximations one finds
\begin{eqnarray}
  \hat{Q}(\kappa) & = & f(\kappa) + \lambda \int_{-\infty}^{\infty} d\nu K(\kappa, \nu) f(\nu) + \\ \nonumber 
    & + & \lambda^2 \int_{-\infty}^{\infty} \int_{-\infty}^{\infty} d\nu d \nu_1 K(\kappa, \nu) K(\nu, \nu_1) f(\nu_1) + \\ \nonumber
      & + & \lambda^2 \int_{-\infty}^{\infty} \int_{-\infty}^{\infty} \int_{-\infty}^{\infty}  d\nu d \nu_1 d\nu_2 K(\kappa, \nu) K(\nu, \nu_1) K(\nu_1,\nu_2) f(\nu_2) + \\ \nonumber
        &\cdots& \\ \nonumber 
\end{eqnarray}

Substituting for our $f(\kappa)$ and $K(\kappa, \nu)$ one finds

\begin{eqnarray}
  \hat{Q}(\kappa) & = & \frac{\frac{1}{2} \Big[C_{kj} \Delta_j\Big]_\kappa} {\frac{m}{2}( \omega^2 - \kappa^2 )} +
  \int_{-\infty}^{\infty} d\nu
  \frac{\frac{1}{2 \pi}\hat{\hat{k}}_{I^{-}}^{k \rightarrow j}(\kappa-\nu,-\nu)}{\frac{m}{2}(\omega^2 - \kappa^2 )}
  \frac{\frac{1}{2} \Big[C_{kj} \Delta_j\Big]_\nu} {\frac{m}{2}( \omega^2 - \nu^2 )}
  + \\ \nonumber 
  & + & \int_{-\infty}^{\infty} \int_{-\infty}^{\infty} d\nu d \nu_1
  \frac{\frac{1}{2 \pi}\hat{\hat{k}}_{I^{-}}^{k \rightarrow j}(\kappa-\nu,-\nu)}{\frac{m}{2}(\omega^2 - \kappa^2 )}
  \frac{\frac{1}{2 \pi}\hat{\hat{k}}_{I^{-}}^{k \rightarrow j}(\nu-\nu_1,-\nu_1)}{\frac{m}{2}(\omega^2 - \nu^2 )}
  \frac{\frac{1}{2} \Big[C_{kj} \Delta_j\Big]_{\nu_1}} {\frac{m}{2}( \omega^2 - \nu_1^2 )}
  \\ \nonumber
  & + & \int_{-\infty}^{\infty} \int_{-\infty}^{\infty} \int_{-\infty}^{\infty}  d\nu d \nu_1 d\nu_2
  \frac{\frac{1}{2 \pi}\hat{\hat{k}}_{I^{-}}^{k \rightarrow j}(\kappa-\nu,-\nu)}{\frac{m}{2}(\omega^2 - \kappa^2 )}
  \frac{\frac{1}{2 \pi}\hat{\hat{k}}_{I^{-}}^{k \rightarrow j}(\nu-\nu_1,-\nu_1)}{\frac{m}{2}(\omega^2 - \nu^2 )}
  \frac{\frac{1}{2 \pi}\hat{\hat{k}}_{I^{-}}^{k \rightarrow j}(\nu_1-\nu_2,-\nu_2)}{\frac{m}{2}(\omega^2 - \nu_1^2 )}
  \frac{\frac{1}{2} \Big[C_{kj} \Delta_j\Big]_{\nu_2}} {\frac{m}{2}( \omega^2 - \nu_2^2 )} + \\ \nonumber
& + \cdots & \\ \nonumber
\end{eqnarray}

The above expression can be
written as $Q(t)=\int_{-\infty}^{\infty} ds \frac{1}{2}G(t,s) C_{kj}(s) \Delta_j (s)$ 
where $G(t,s)$ stands for the  iterated sum 
\begin{eqnarray}
G(t,s) & = & \Bigg[ \frac{1}{ 2\pi} \int_{-\infty}^{\infty} d\kappa e^{-i \kappa t} 
  \frac{1}{\frac{m}{2}( \omega^2 - \kappa^2 )} e^{i \kappa s}  
  +  \\ \nonumber
  & + & \frac{1}{2 \pi} \int_{-\infty}^{\infty} d\kappa e^{-i \kappa t} \frac{1}{\frac{m}{2}( \omega^2 - \kappa^2 )}
  \int_{-\infty}^{\infty} d\nu \frac{ \frac{1}{2 \pi} \hat{\hat{k}}_{I^{-}}^{k \rightarrow j}(\kappa-\nu,-\nu)}{\frac{m}{2}(\omega^2 - \nu^2 )}
  e^{i \nu s} + \\ \nonumber 
  & + & \frac{1}{2 \pi} \int_{-\infty}^{\infty} d\kappa e^{-i \kappa t} \frac{1}{\frac{m}{2}( \omega^2 - \kappa^2 )}
  \int_{-\infty}^{\infty} \int_{-\infty}^{\infty} d\nu d \nu_1
  \frac{\frac{1}{2 \pi}\hat{\hat{k}}_{I^{-}}^{k \rightarrow j}(\kappa-\nu,-\nu)}{\frac{m}{2}(\omega^2 - \nu^2 )}
  \frac{\frac{1}{2 \pi}\hat{\hat{k}}_{I^{-}}^{k \rightarrow j}(\nu-\nu_1,-\nu_1)}{\frac{m}{2}(\omega^2 - \nu_1^2 )}
   e^{i \nu_1 s}  + \cdots \Bigg]  \\ \nonumber
\end{eqnarray}
The first (zero order) term in the sum
\begin{equation}
G_0(t,s) = \frac{1}{ 2 \pi} \int_{-\infty}^{\infty} d\kappa e^{i \kappa (s-t)} 
  \frac{1}{\frac{m}{2}( \omega^2 - \kappa^2 )}
\end{equation}
When $\kappa$ has infinitesimal negative imaginary part
and when $s-t$ is negative, the integral can be closed in the lower half plane, and is zero. When $s-t$ is positive the integral can be closed in the upper complex plane and is $\frac{2}{m\omega}\sin\omega (s-t)$.

The next (first order) term \textbf{is}
\begin{eqnarray}
G_1(t,s) & = & \frac{1}{2 \pi} \int_{-\infty}^{\infty} d\kappa e^{-i \kappa t} \frac{1}{\frac{m}{2}( \omega^2 - \kappa^2 )}
  \int_{-\infty}^{\infty} d\nu  \frac{e^{i \nu s}}{\frac{m}{2}(\omega^2 - \nu^2 )} \frac{1}{2 \pi} \hat{\hat{k}}_{I^{-}}^{k \rightarrow j}(\kappa-\nu,-\nu) = \\ \nonumber 
  & = & \frac{1}{2 \pi} \int_{-\infty}^{\infty} d\kappa \frac{e^{-i \kappa t}}{\frac{m}{2}( \omega^2 - \kappa^2 )}
  \int_{-\infty}^{\infty} d\nu  \frac{e^{i \nu s}}{\frac{m}{2}(\omega^2 - \nu^2 )} \frac{1}{2 \pi} \int_{-\infty}^{\infty} dt_1 e^{i (\kappa-\nu) t_1}
  \hat{k}_{I^{-}}^{k \rightarrow j}(t_1,-\nu) = \\ \nonumber
  & = & \frac{1}{2 \pi} \int_{-\infty}^{\infty} d\kappa \frac{e^{-i \kappa t}}{\frac{m}{2}( \omega^2 - \kappa^2 )}
  \int_{-\infty}^{\infty} d\nu  \frac{e^{i \nu s}}{\frac{m}{2}(\omega^2 - \nu^2 )} \frac{1}{2 \pi} \int_{-\infty}^{\infty} dt_1 e^{i (\kappa-\nu) t_1}
  \int_{-\infty}^{\infty} dt_2 e^{- i \nu (t_2-t_1) } k_{I}^{k \rightarrow j}(t_1,t_2-t_1) \Theta(t_2-t_1)
\end{eqnarray}
\noindent where the definitions of $\hat{\hat{k}}_{I^{-}}^{k \rightarrow j}(\kappa-\nu,-\nu)$ and $\hat{k}_{I}^{k \rightarrow j}(t,-\nu)$ given in Eqs. (\ref{eq:kidhat}) and (\ref{eq:kidhat}) were used. In the above some term can be rearranged to give
\begin{eqnarray}
  G_1(t,s)   & = &  \frac{1}{2 \pi} \int_{-\infty}^{\infty} dt_1 
  \int_{-\infty}^{\infty} dt_2 \frac{1}{2 \pi} k_{I}^{k \rightarrow j}(t_1,t_2-t_1) \Theta(t_2-t_1)
  \int_{-\infty}^{\infty} d\kappa \frac{e^{-i \kappa (t-t_1)}}{\frac{m}{2}( \omega^2 - \kappa^2 )}
  \int_{-\infty}^{\infty} d\nu  \frac{e^{i \nu (s-t_2) }}{\frac{m}{2}(\omega^2 - \nu^2 )} = \\ \nonumber
  & = & 
  \int_{-\infty}^{\infty} dt_1 \int_{t_1}^{\infty} dt_2 k_{I}^{k \rightarrow j}(t_1,t_2-t_1) 
  \frac{1}{2 \pi} \int_{-\infty}^{\infty} d\kappa \frac{e^{-i \kappa (t-t_1)}}{\frac{m}{2}( \omega^2 - \kappa^2 )}
   \frac{1}{2 \pi} \int_{-\infty}^{\infty} d\nu  \frac{e^{i \nu (s-t_2) }}{\frac{m}{2}(\omega^2 - \nu^2 )} =
   \\ \nonumber
   & = &
   \int_{-\infty}^{\infty} dt_1 \int_{t_1}^{\infty} dt_2 k_{I}^{k \rightarrow j}(t_1,t_2-t_1) G_0(t_1-t) G_0(s-t_2) 
\end{eqnarray}
Successive repetition of the above procedure leads  to the following expression
\begin{equation}
G(t,s-t) =  G_0(s-t)+ \int_{t}^{T} dt_1 \int_{t_1}^{T} dt_2 G_0(t_1-t) k_{I}^{k \rightarrow j}(t_1,t_2-t_1) G(t_2,s-t_2),
\end{equation}
which is Eq \eqref{eq:G-twinning}. 
In general, the above one-sided functional
equation does not have a 
convenient closed-form solution.
However, if one assumes 
that both 
$k_{I}^{k \rightarrow j}(t,s-t)$ and
$G(t,s-t) $ depend only on their
second argument and essentially vanish
when it is large enough,
one has the considerably
simpler relation
\begin{equation}
G(s-t) =  G_0(s-t)+ \int_{t}^{\infty} dt_1 \int_{t_1}^{\infty} dt_2 G_0(t_1-t) k_{I}^{k \rightarrow j}(t_2-t_1) G(s-t_2) 
\end{equation}
valid when $s$ (the largest time
in above) is considerably smaller
than $T$ (the final time).
Since this equation involves a convolution, it can be conveniently written in the Laplace domain as
\begin{equation}
\tilde{G}(\lambda) =  \tilde{G}_0(\lambda)+\tilde{G}_0(\lambda) k_{I}^{k \rightarrow j}(\lambda) \tilde{G}(\lambda),
\label{eq:G-Laplace-twinning}
\end{equation}
where for simplicity we used the same symbol for $k_{I}^{k \rightarrow j}$  in the time domain and in the Laplace domain.
The Laplace transform $\tilde{G}_0(\lambda)$ is
given by
\begin{eqnarray}
\tilde{G}_0(\lambda) &=& \int_0^{\infty}e^{-\lambda t}  
\frac{2}{m\omega}\sin\omega t\, dt 
= \frac{2}{m} \frac{1}{\lambda^2 + \omega^2}
\label{eq:G0-Laplace}
\end{eqnarray}
This is twice the response function of the
harmonic oscillator as conventionally defined.

\section{Fixed point for constant interactions on $n$-regular random graphs} 
\label{sec:fixed-point}
The goal of this section is to 
make an explicit computation of (\ref{eq:messpassings}) in a specific model.
We will present two approaches, one relying on the known representation
of the Laplace transform function,
and one of a sine transform analogous to the the
Mehler–Sonine representation of the Bessel function.
We consider a set of oscillators interacting with constant couplings,
what one may call a ferromagnetic case.
In the present context constant interaction means that all functions $C_{ij}(t)$ are the same and do not depend on time.
Moreover, we will assume that oscillators are placed in an $n$-regular random graphs, i.e. a graph where all vertices have the same number ($n$) of neighbours. In such a setting 
there can be a deterministic Replica Symmetric phase  of the cavity equations where all cavity messages are the same. In this section we will consider the corresponding fixed point and the corresponding message.

In this case, it is easy to show that:
\begin{eqnarray}
  &\tilde{k}_I^{k \rightarrow j}(t,s-t)  & =  \frac{1}{2} C_{kj} C_{kj} G^{k \rightarrow j}(t,s-t) \\
  & \tilde{k}_R^{k \rightarrow j}(t,s) &   =  C_{kj} C_{kj} \int_{-\infty}^t \int_{-\infty}^s k_R^{k \rightarrow j}(t',s') G^{k \rightarrow j}(t',t-t') G^{k \rightarrow j}(s',s-s')
  \label{eq:kIkR}
\end{eqnarray}
These expression apparently suggest that the sign of the couplings $C_{ij}$ is irrelevant, they always appear squared. 
However, we have to remember that in above we have implicitly used the relation $\omega^2 = \omega_0^2 + \frac{nC}{m}$
which enters in the bare response function $G_0$, which in turn enters in $G$. $C$ can be taken arbitrarily large positive,
but not smaller than $-\frac{m\omega_0^2}{n}$, as otherwise the total potential is not positive definite and the system has no ground state. We now turn to the two different ways in which the fixed point of the imaginary kernel can be derived.

\subsection{First version of the calculation}
\label{sec:Mulet-calculation}
We start by summing the twinning relation
\eqref{eq:G-Laplace-twinning}
which gives
\begin{equation}
\tilde{G}(\lambda) =  \tilde{G}_0(\lambda)+\tilde{G}_0(\lambda) k_{I}^{k \rightarrow j}(\lambda) \tilde{G}(\lambda) = \tilde{G}_0(\lambda)\Big[ 1-\tilde{G}_0(\lambda) k_{I}^{k \rightarrow j}(\lambda)\Big]^{-1} 
\end{equation}

The Laplace transform of the kernels (Eq. (\ref{eq:kIkR})) gives

\begin{eqnarray}
  &\tilde{k}_I^{k \rightarrow j}(\lambda)  & =  \frac{1}{2} C_{kj} C_{kj} \tilde{G}^{k \rightarrow j}(\lambda) \nonumber \\
 & \tilde{k}_R^{k \rightarrow j}(\lambda) &   =  C_{kj} C_{kj} k_R^{k \rightarrow j}(\lambda) (\tilde{G}^{k \rightarrow j})^2(\lambda).
\end{eqnarray}
Now we use the fact that $F_{j \rightarrow i} = \sum_k \tilde{F}_{k \rightarrow j} $ as well as $k_I= (n-1) \tilde{k}_I$ and $k_R= (n-1) \tilde{k}_R$, and we assume that all the couplings are the same. Therefore:

\begin{eqnarray}
\label{eq:derv_Rob}
& k_I(\lambda)= (n-1) \tilde{k}_I(\lambda)& = (n-1) \frac{1}{2} C^2 \tilde{G}(\lambda) \nonumber \\
& k_R(\lambda) = (n-1) \tilde{k}_R(\lambda)& = (n-1) C^2 k_R \tilde{G}^2(\lambda),
\end{eqnarray}
Subsequently we use definition of $G(\lambda)$ to solve Eq. (\ref{eq:derv_Rob}) for $k_I(\lambda)$ and find
\begin{eqnarray}
k_I(\lambda) = & (n-1) \frac{1}{2} C^2  \tilde{G}_0(\lambda)\Big[ 1-\tilde{G}_0(\lambda) k_{I}^{k \rightarrow j}(\lambda)\Big]^{-1}  
= &\frac{\tilde{G}_0^{-1}(\lambda)}{2} \Bigg[ 1 \pm \sqrt{1- 2 (n-1) C^2 \tilde{G}_0^2(\lambda)}\Bigg]
\end{eqnarray}
From \eqref{eq:G0-Laplace} we know that
$\tilde{G}_0(\lambda) = \frac{2}{m} \frac{1}{\lambda^2 + \omega^2}$. It is clear we should take the negative 
sign in front of the square root, as otherwise the Laplace 
transform does not decay with parameter $\lambda$ at infinity.
To derive the actual message as a function of time we follow the definition of $ k_{I}^{k \rightarrow j}(\lambda)$,

\begin{eqnarray}
  k_{I}(\lambda) & = & 2(n-1) \frac{C^2}{m}
  \frac{1}{  (\omega^2+\lambda^2) +  \sqrt{ (\omega^2+\lambda^2)^2 - (\frac{2}{m})^2 2 (n-1)C^2 }}  \\ \nonumber
   &=& \frac{m}{4}
  a^4
   \frac{1}{ (\omega^2+\lambda^2) + \sqrt{ (\lambda^2 +\omega_1^2) (\lambda^2 + \omega_2^2) }}
\label{eq:kIlambda}
\end{eqnarray}

\noindent where $a^4 = (\frac{8 (n-1)}{m^2}) C^2$, and $\omega_1^2 = \omega^2-a^2$ and $\omega_2^2= \omega^2+a^2$. 

This expression (\ref{eq:kIlambda}) can be rewritten in a convenient form

\begin{equation}
  k_{I}(\lambda) = \frac{m}{4}
\Bigg\{
  \frac{a^4}{\sqrt{ (\lambda^2+\omega_1^2) (\lambda^2 + \omega_2^2) }} - \Bigg[ \frac{ (\lambda^2 + \omega^2)^2}{\sqrt{ (\lambda^2+\omega_1^2) (\lambda^2 + \omega_2^2) }} -( \lambda^2+\omega^2) \Bigg] \Bigg\}
\end{equation}
The inverse Laplace transform of the first term is \cite{GR}:

\begin{equation}
  \mathcal{L}^{-1}[\frac{1}{\sqrt{ (\lambda^2+\omega_1^2) (\lambda^2 + \omega_2^2) }}]
  =  \int_0^t d\tau J_0(\omega_1 \tau)  J_0(\omega_2 (t-\tau) ) \equiv f(t) 
\end{equation}

To compute the inverse transform of the terms in bracket we exploit the property of the Laplace transform 
$
  \mathcal{L}[f^{(n)}(t)] = \lambda^n F(\lambda) - \sum_{k=1}^{n} \lambda^{n-k} f^{k-1}(0)
$,
where $f^k(0)$ denotes $k$-th derivative of $f(t)$ calculated $t=0$. In the case considered here we have
$ \mathcal{L}^{-1}[\lambda^4 F(\lambda)] = f^{(4)}(t) + \mathcal{L}^{-1}[ \sum_{k=1}^4\lambda^{4-k} f^{(k-1)}(0)]$, and
$  \mathcal{L}^{-1}[\lambda^2 F(\lambda)] = f^{(2)}(t) + \mathcal{L}^{-1}[ \sum_{k=1}^2 \lambda^{2-k} f^{k-1}(0)]$, where $F(\lambda) = \frac{1}{\sqrt{ (\lambda^2+\omega_1^2) (\lambda^2 + \omega_2^2) }}$ .
Taking those relations into account one arrives at the expression for the imaginary kernel in the time domain
\begin{equation}
  k_{I}(\tau) =  \mathcal{L}^{-1}[k_{I}(\lambda)] = 
\frac{m}{4} \Bigg\{a^4 f(\tau) - \Big[  f^{(4)}(\tau) + 2 \omega^2  f^{(2)}(\tau) + \omega^4 f(\tau)\Big] \Bigg\},
\end{equation}
where in derivation we the fact that $f(0)=0$, $f^{(1)}(0)=1$, $f^{(2)}(0)=0$,$f^{(3)}(0)=-\omega^2$, $f^{(4)}(0)=0$.
In this way we have reduced the 
inverse Laplace transform to a convolution of
Bessel functions and derived combinations
thereof.

\subsection{Second version of the calculation}
\label{sec:Erik-Jan-calculation}
For an alternative version of the calculation it is convenient
to restate the iteration of the Vernon transform in the Laplace domain
as
\begin{equation}
\label{eq:iteration-equation}
\tilde{k}_I^{n+1}(\lambda)
= \frac{(n-1)C^2}{2}
\tilde{G}_0(\lambda)
\left(1- \tilde{G}_0(\lambda)
\tilde{k}_I^{n}(\lambda)\right)^{-1}
\end{equation}
where as above
$\tilde{G}_0(\lambda)=\frac{2}{m}\frac{1}{\lambda^2+\omega^2}$ is 
twice the
Laplace transform of the
harmonic oscillator response function.

The fixed point can thus be written
(equivalent to \eqref{eq:kIlambda})
as
\begin{eqnarray}
\label{eq:k-solution-2}
\tilde{k}_I^*(\lambda) 
&=& m\frac{\lambda^2+\omega^2}{4}
\left(1-\sqrt{1-\frac{8(n-1)\, C^2 }{m^2(\lambda^2 +\omega^2)^2}}\right)  \end{eqnarray}
The fixed point kernel in the time domain is given by an inverse 
Laplace transform:
\begin{eqnarray}
\label{eq:inverseLaplace}
k_I(\tau) = \frac{1}{2\pi i} \int \tilde{k}_I(\lambda) e^{\lambda\tau} d\lambda
\end{eqnarray}
The integral is to be performed on a vertical contour far enough
to the right in the complex plane. Since the integrand goes down
as $\lambda^{-2}$ for large $\lambda$ such an integral has a
finite value. If the contour can be moved to the far left in the 
complex plane, then that integral will be zero because $e^{\lambda\tau}$
then acts as dampening. The inverse Laplace transform is hence given by
the integrals encircling poles and cut-lines encountered when moving
the integral contour
as illustrated in Fig.~\ref{fig:c}.

\begin{figure}
\centering
\includegraphics[scale=0.1]{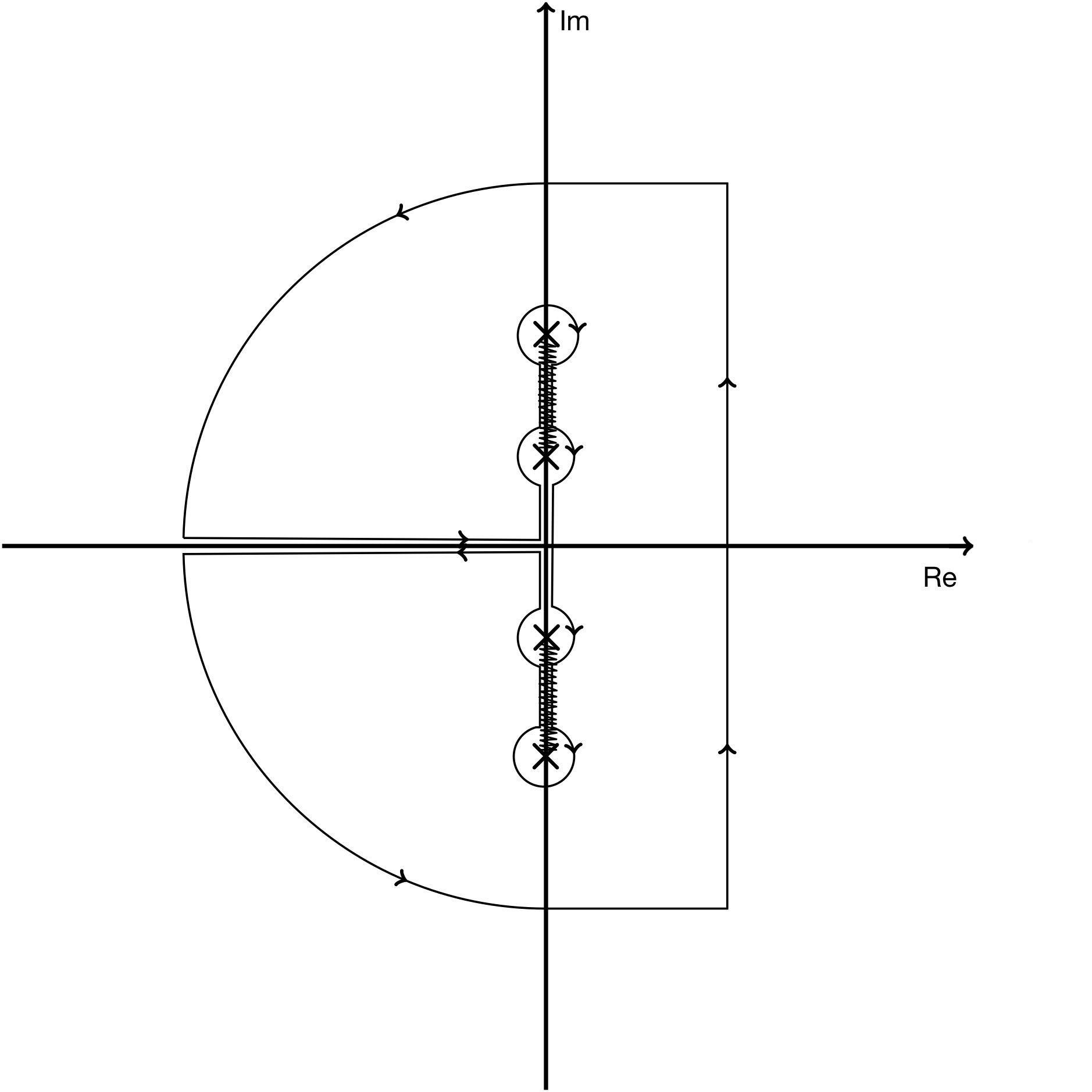}
		\caption{\label{fig:c}
		Illustration of the 
		computation of the inverse
		Laplace transform of the fixed point kernel from Eq.~(\protect\ref{eq:inverseLaplace}). By definition this is the integral over a vertical path far enough to the right in the complex plane. Moving the path to the left part of the complex plane one picks up branch cut contributions determined in the text}
\end{figure}

The kernel $\tilde{k}_I(\lambda)$ is analytic everywhere except
in the neighbourhood of points where the argument of the
square root vanishes. These points are at
\begin{equation}
i \lambda_{\pm\pm} = \pm i \sqrt{\omega^2 \pm \sqrt{8(n-1)}C/m}
\end{equation}
The arguments of the outer square
root in above is positive, hence the four points all lie on the imaginary
axis. The kernel $\tilde{k}_I(\lambda)$ is analytic around the real line
as well as for large enough $\lambda$. The contour therefore needs to
encircle two cuts between respectively $i(\lambda_{+-},\lambda_{++})$
and $i(\lambda_{--},\lambda_{-+})=i(-\lambda_{++},-\lambda_{+-})$.

It is convenient to re-write the square root 
in \eqref{eq:k-solution-2} as $ \sqrt{(\lambda^2+\lambda^2_{+-})(\lambda^2+\lambda^2_{++})}$. On the imaginary axis the argument of the square root is then
positive for $z<\lambda_{+-}$, negative
in the interval $(\lambda_{+-},\lambda_{++})$, and positive
again for $z>\lambda_{++}$.
The phase of the square root is zero on the 
imaginary axis up to just below the start of the cut at $i\lambda_{+-}$.
Along the cut and just to the right the absolute value of the
square root is
$ \sqrt{(z^2-\lambda^2_{+-})(-z^2+\lambda^2_{++})}$
and the phase is $i$.
At the same point along the cut 
and just to the left the phase is $-i$.
The value of the integral encircling $(\lambda_{+-},\lambda_{++})$
in the counter-clockwise direction is hence 
\begin{eqnarray}
\hbox{($+$)-side} &=& 
\int_{\lambda_{+-}}^{\lambda_{++}}
 e^{iz\tau}\, (2i) (-\frac{1}{2})
 \sqrt{(z^2-\lambda^2_{+-})(-z^2+\lambda^2_{++})},  d(iz) 
\end{eqnarray}

where $-\frac{1}{2}$ is the pre-factor of the square root in
\eqref{eq:k-solution-2}.

For the integral encircling $(\lambda_{--},\lambda_{-+})$
one can start from that the phase of the square root must be
zero on the imaginary axis just above the cut.
Along the cut and just to the left the phase is $+i$,
and to the right it is $-i$.
The value of this integral, encircling this cut in the positive
direction, is thus

\begin{eqnarray}
\hbox{($-$)-side} &=& 
\int_{-\lambda_{++}}^{-\lambda_{+-}}
 e^{iz\tau}\, (-2i) (-\frac{1}{2})
 \sqrt{(z^2-\lambda^2_{+-})(-z^2+\lambda^2_{++})},  d(iz) 
\nonumber \\
&=& 
\int_{\lambda_{+-}}^{\lambda_{++}}
 e^{-iz\tau}\, (-2i) (-\frac{1}{2})
 \sqrt{(z^2-\lambda^2_{+-})(-z^2+\lambda^2_{++})},  d(iz) 
\end{eqnarray}

Combining both integrals 
and bringing out a 
dimensional factors
we have a rather simple 
integral representation
\begin{equation}
\label{eq:inverseLaplace-2}
k_I(\tau) = 
\Lambda
\int_{q}^{1}
\sin (\lambda_{++}x\tau) 
\sqrt{(x^2-q^2)(1-x^2)}\, dx 
\end{equation}
where 
$\Lambda =m
\lambda^3_{++}/\pi$,
and where 
we have used $q=\frac{\lambda_{+-}}{\lambda_{++}}$.
The expression in \eqref{eq:inverseLaplace-2}
is analogous to the 
Mehler–Sonine representation of the Bessel function $J_0(\omega t)$, which is in fact
nothing but the inverse Laplace transform of the function
$\frac{1}{\sqrt{\Lambda^2 + \omega^2}}$.

The representation  \eqref{eq:inverseLaplace-2} lends itself to
a physical interpretation as follows.
The total Hamiltonian 
in the tree subtended from $j$
will have normal modes. The
kernel of the real
Feynman-Vernon action on $i$ 
is according to the general 
formula
\begin{equation}
\label{eq:equivalent-spectrum}
k_I(\tau) = \int 
d\omega \sin\omega\tau J(\omega)
\, d\omega
\end{equation}
where $J(\omega)$ is the spectral density.
Comparing 
\eqref{eq:inverseLaplace-2}
and \eqref{eq:equivalent-spectrum}
we have the non-trivial result
\begin{equation}
J(\omega) = \left\{
\begin{array}{ll}
\sqrt{(\omega^2-\lambda_{+-}^2)(\lambda_{++}^2-\omega^2)} &
\hbox{for $\omega\in [\lambda_{++},\lambda_{+-}]$}\\
0 & \hbox{otherwise}\\
\end{array}\right. \nonumber
\end{equation}
In other words, the infinite network
as to its influence on one
system, behaves for this 
ferromagnetic harmonic oscillator
example as a bath 
with compact spectral support.

The Fourier transform of the function 
$k_I(\tau)$ in \eqref{eq:equivalent-spectrum},
when the value is zero for negative $\tau$,
is the Laplace transform of $k_I(\tau)$  
of argument $\i\nu$. 
Since $\hat{k}_I(-\nu)=\left(\hat{k}_I(\nu)\right)^*$
it is enough to consider positive $\nu$.
The function $k_I(i\nu)$ is hence real except on the
cut-line (on positive imaginary $\lambda$ axis) where it is
\begin{eqnarray}
\label{eq:k-solution-3}
\hat{k}_I^*(\nu) 
&=& m\frac{-\nu^2+\omega^2}{4}
\left(1-i\sqrt{\frac{4(n-1)\, C^2 }{m^2(-\nu^2 +\omega^2)^2}-1}\right) \nonumber 
\end{eqnarray}
The Fourier transform of a real function 
satisfies $\hat{k}_I^*(-\nu)=(\hat{k}_I^*(\nu))^*$.
We can th
We note for further reference that
on the cut-lines we have 
\begin{eqnarray}
\label{eq:k-solution-3}
\hat{k}_I^*(\nu) \hat{k}_I^*(-\nu) 
&=& \left(m\frac{-\nu^2+\omega^2}{4}\right)^2
\frac{8(n-1)\, C^2 }{m^2(-\nu^2 +\omega^2)^2}
= \frac{(n-1)\, C^2 }{2} 
\end{eqnarray}
Note that the cavity kernels 
$\hat{k}_I^*(\nu)$ are of the Belief Propagation $n$-type
messages
(variables to interactions).

\section{Iteration of the real kernel}
In  this section we analyze the coefficient of the linear transformation $\mathcal{W}$ in the uniform network. It is convenient to do this in the Fourier domain. From Eq. (\ref{eq:kIkR}) one finds
\begin{eqnarray}
\hat{k}_R^{j \rightarrow i}(\nu) &=&
\frac{4}{C^2}| \hat{k}_I^{j \rightarrow i}(\nu)|^2 k_R^{j \rightarrow i \; *}(\nu) (n-1)
\end{eqnarray}
The cavity kernels $k_I$ that appear here are of
the  Belief Propagation $m$-type
(interactions to variables). At the uniform
fixed point they are $(n-1)^{-1}$ times the
 Belief Propagation $n$-type
messages which appear in \eqref{eq:k-solution-3}.
Combining everything we have the rather simple result
\begin{eqnarray}
\frac{4}{C^2}| \hat{k}_I^{j \rightarrow i}(\nu)|^2 = 2
\end{eqnarray}
Asymptotically all Fourier components of $k_R$ corresponding
to the spectrum of the equivalent environment
grow under the iteration (multiplier is two).
On the other hand it is easy to see that other Fourier
components sufficiently far from the spectrum decay (multiplier less than one).

\begin{figure*}
\centering
\includegraphics[scale=0.5]{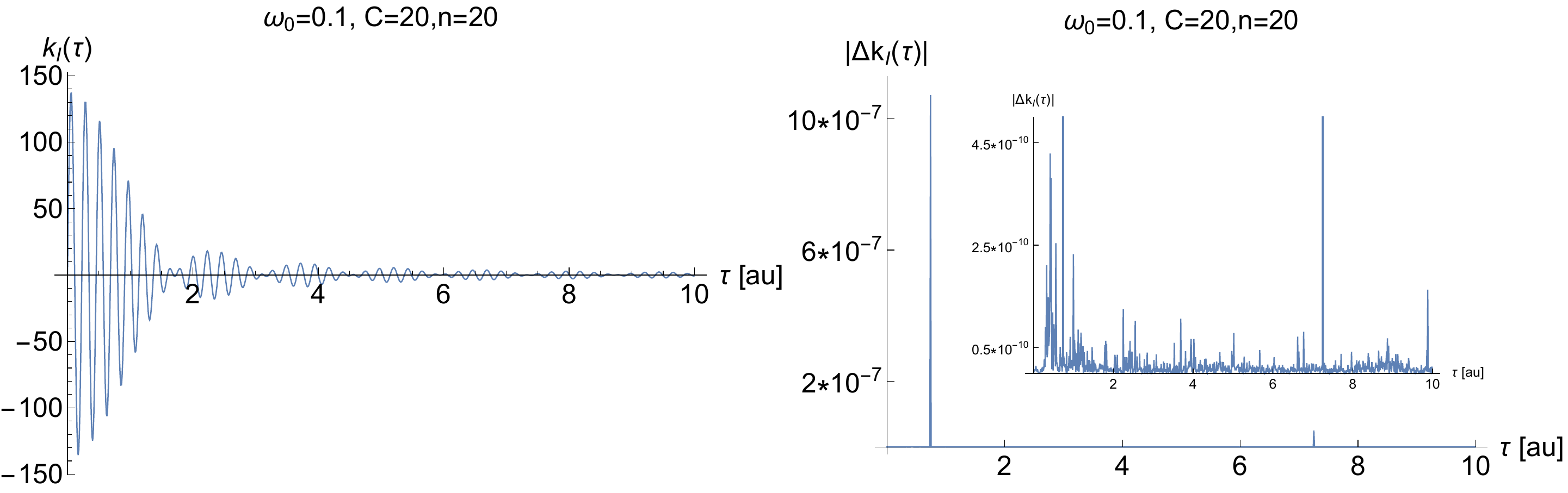}
		\caption{ Numerical comparison of the two formulas for the inverse Laplace transform of the imaginary kernel Eq. (\ref{eq:inverseLaplace}) and Eq. (\ref{eq:inverseLaplace-2}). In the left panel the two formulas are plotted together. In the right panel absolute value of the difference between the two formulas (note a different scale in the inset).
		The parameters used to compute the kernels are The parameters are: $\omega_0=0.1$, $C=20$, $n=20$, $m=\frac{1}{2}$.. Extensive numerical comparison of the two formulas for different set of parameters allows to conclude that they are the same.         \label{fig:Comparison}
}
\end{figure*}

\section{Stability Analysis of the deterministic Replica Symmetric solution
for ferromagnetic model on $n$-regular random graphs}
\label{sec:stability}
In this section we
consider the stability of the fixed 
point found in Section II of the main paper,
in the context of Replica Symmetry.
We hence here consider parameters such that the
fixed point exists.
The starting point is then that 
$k$ and $\tilde{k}$ in that analysis 
are not values but arguments of probability distributions
that satisfy compatibility conditions.
The goal is to check whether Dirac delta distributions
are stable solutions of these compatibility conditions.
We thus start from the BP update equations for
Feynman-Vernon kernels in Laplace transform picture
written as

\begin{equation}
\label{eq:q-iteration}
Q(k) = \int \prod_k^{n-1} d\hat{k}_j P( \hat{k}_j) \delta( k - \sum_j \hat{k}_j )
\end{equation}
and

\begin{equation}
\label{eq:p-iteration}
P(\hat{k}) = \int dk  Q(k) \delta( \hat{k} - f_{\lambda}(k) )
\end{equation}
where $ f_{\lambda}(k) = \frac{C^2}{2} \frac{ G_0(\lambda)}{ 1-  G_0(\lambda) k}\big]$ to be expanded around $k=k_0= (n-1) k^*$ is the Vernon
transform applied to Laplace transform variable with parameter $\lambda$.
$G_0(\lambda)=\frac{3}{m}\frac{1}{\lambda^2+\omega^2}$
is twice the response function of the free harmonic oscillator.

Instead of taking $P$ and $Q$ delta functions we then assume
$P(\hat{k}) \sim e^{- \frac{1}{ 2 \sigma^2} ( \hat{k}- k^* )^2}$
That from \eqref{eq:q-iteration} leads directly to:
\begin{equation}
Q(k)  \sim e^{- \frac{1}{ 2 \sigma^2 (n-1)} ( k- (n-1) k^* )^2}
\end{equation}

The point now is to check, whether this is consistent with a {\em new} Gaussian $P(\hat{k})$. This is clearly not the case for general $ f_{\lambda}(k)$. Therefore a reasonable approach is to check if for $\sigma$ small enough, the variance of $P(\hat{k})$ grows or goes to zero. In the second case, we say that the 
deterministic solution $k^*$
\eqref{eq:q-iteration} and
\eqref{eq:p-iteration}
is stable. Otherwise, it is not.

We then proceed to estimate $ \langle \hat{k}^\alpha \rangle$, which in practice translates into solving the following integral:
\begin{eqnarray}
  \langle \hat{k}^\alpha \rangle & = & \int d\hat{k}^\alpha \int dk Q(k) \delta( \hat{k} - f_{\lambda}(k) ) = \nonumber \\
  \int dk Q(k) f_{\lambda}(k)^\alpha &=& \int dk  e^{- \frac{1}{ 2 \sigma^2 (n-1)} (k- (n-1) k^* )^2} f_{\lambda}(k)^\alpha \nonumber
\end{eqnarray}
We expand the function $ f_{\lambda}(k)^\alpha = \big[ \frac{C^2}{2} \frac{ G_0(\lambda)}{ 1-  G_0(\lambda) k}\big]^\alpha$ around $k=k_0= (n-1) k^*$.
For the expected value
($\alpha=1$)  we have

\begin{eqnarray}
  \frac{ G_0(\lambda)}{ 1-  G_0(\lambda) k} & = & \frac{ G_0(\lambda)}{ 1-  G_0(\lambda) k_0} + \big[ \frac{ G_0(\lambda)}{ 1-  G_0(\lambda) k_0}\big]^2(k-k_0) +\big[\frac{ G_0(\lambda)}{ 1-  G_0(\lambda) k_0}\big]^3(k-k_0)^2 
\end{eqnarray}

such that

\begin{eqnarray}
  \langle k \rangle & = & \frac{C^2}{2}
  \Big[ \frac{ G_0(\lambda)}{ 1-  G_0(\lambda) k_0} \int dk e^{-\frac{1}{2 \sigma^2(n-1)} (k-k_0)^2} + \big(\frac{ G_0(\lambda)}{ 1-  G_0(\lambda) k_0}\big)^3\int dk e^{-\frac{1}{2 \sigma^2(n-1)} (k-k_0)^2} (k-k_0)^2 \Big] = \\ \nonumber
    &= & \frac{C^2}{2}\frac{ G_0(\lambda)}{ 1-  G_0(\lambda) k_0} \Big[1 + \sigma^2(n-1) \big(\frac{ G_0(\lambda)}{ 1-  G_0(\lambda) k_0}\big)^2 \Big]
\end{eqnarray}

Similarly, for the second moment ($\alpha=2$):

\begin{eqnarray}
  \left[ \frac{ G_0(\lambda)}{ 1-  G_0(\lambda) k}\right]^2 & = &
  \left[\frac{ G_0(\lambda)}{ 1-  G_0(\lambda) k_0}\right]^2 + 2 \left[ \frac{ G_0(\lambda)}{ 1-  G_0(\lambda) k_0}\right]^3(k-k_0) + \frac{6}{2} \left[\frac{ G_0(\lambda)}{ 1-  G_0(\lambda) k_0}\right]^4 (k-k_0)^2   
\end{eqnarray}

and

\begin{eqnarray}
  \langle k^2 \rangle & = & \frac{C^4}{4} \big[\frac{ G_0(\lambda)}{ 1-  G_0(\lambda) k_0}\big]^2  \Big[ 1+ 3 \sigma^2(n-1) \big[\frac{ G_0(\lambda)}{ 1-  G_0(\lambda) k_0}\big]^2 \Big]
\end{eqnarray}

Putting everything together we have

\begin{eqnarray}
  \langle k^2 \rangle - \langle k \rangle^2 & = & \sigma^2 \frac{(n-1)}{4} \nonumber \\
  &&\quad C^4
  \big[\frac{ G_0(\lambda)}{ 1-  G_0(\lambda) k_0}\big]^4 
  \label{eq:iterated variance}
\end{eqnarray}
The variance of the distribution after passing through the BP update
is hence proportional to original variance $\sigma^2$.
Note that the combination 
$C\frac{ G_0(\lambda)}{ 1-  G_0(\lambda) k_0}$
is dimension-less,
and that $\sigma$ has the same dimension as $k$.

Since we expand around the fixed point $k^*$ we have $k_0= (n-1)k^*$,
$\frac{ G_0(\lambda)}{ 1-  G_0(\lambda) k_0}
= \frac{2}{C^2(n-1)}k^*$
and $k^* = m\frac{\lambda^2+\omega^2}{4}
\left(1-\sqrt{1-\frac{8(n-1)\, C^2 }{m^2(\lambda^2 +\omega^2)^2}}\right)$.
For small $C$ we can neglect the difference between $\omega$
and $\omega_0$ and have 
\begin{equation}
    k^* \approx  \frac{(n-1)\, C^2 }{m(\lambda^2 +\omega_0^2)}\quad\hbox{($C$ small)}
\end{equation}
and hence 
\begin{equation}
    \frac{ G_0(\lambda)}{ 1-  G_0(\lambda) k_0} \approx  \frac{1}{(\lambda^2 +\omega_0^2)}\quad\hbox{($C$ small)}
\end{equation}
For small values of $C$ the proportionality is hence less than
one, and by iteration the Gaussian gets sharper. This means that the deterministic (delta-function) solution is stable.


\end{widetext}

\bibliography{references}

\end{document}